\documentclass[notitlepage,letterpaper,aps,onecolumn,amsmath,amsfonts,nofootinbib,preprintnumbers, superscriptaddress,eqsecnum,secnumarabic]{revtex4-1}
\usepackage{graphicx,slashed}
\pdfoutput=1
\usepackage{amssymb,amsmath,latexsym,mathrsfs}
\usepackage[usenames,dvipsnames]{color}
\usepackage{graphicx} 
\usepackage{bm}
\usepackage{natbib}
\usepackage{todonotes}
\usepackage{color}
\usepackage[normalem]{ulem}
\usepackage[breaklinks,colorlinks,urlcolor=blue,citecolor=blue,linkcolor=blue]{hyperref}
\usepackage{epsfig,graphicx,verbatim,xspace,multirow,mathtools}
\usepackage{array}

\def\gsim{\mathrel{\raise.3ex\hbox{$>$\kern-.75em\lower1ex\hbox{$\sim$}}}}

\newcommand{\Eq}[1]{Eq.~\eqref{#1}}
\newcommand{\Fig}[1]{Fig.~\ref{#1}}
\newcommand{\Tab}[1]{Tab.~\ref{#1}}
\newcommand{\Sec}[1]{Section~\ref{#1}}

\begin{document}

\preprint{\tt IFIC/18-15}

\title{Constraining the primordial black hole abundance with 21cm cosmology}

\author{Olga Mena} 
\author{Sergio Palomares-Ruiz}
\author{Pablo Villanueva-Domingo}
\author{Samuel J.~Witte}
\affiliation{Instituto de F\'{\i}sica Corpuscular (IFIC), CSIC-Universitat de Val\`encia, \\
Apartado de Correos 22085,  E-46071, Spain}

\begin{abstract}
The discoveries of a number of binary black hole mergers by LIGO and VIRGO has reinvigorated the interest that primordial black holes (PBHs) of tens of solar masses could contribute non-negligibly to the dark matter energy density. Should even a small population of PBHs with masses $\gtrsim \mathcal{O}(M_\odot)$ exist, they could profoundly impact the properties of the intergalactic medium and provide insight into novel processes at work in the early Universe. We demonstrate here that observations of the 21cm transition in neutral hydrogen during the epochs of reionization and cosmic dawn will likely provide one of the most stringent tests of solar mass PBHs. In the context of 21cm cosmology, PBHs give rise to three distinct observable effects: {\emph{(i)}} the modification to the primordial power spectrum (and thus also the halo mass function) induced by Poisson noise, {\emph{(ii)}} a uniform heating and ionization of the intergalactic medium via X-rays produced during accretion, and {\emph{(iii)}} a local modification to the temperature and density of the ambient medium surrounding isolated PBHs. Using a four-parameter astrophysical model, we show that experiments like SKA and HERA could potentially improve upon existing constraints derived using observations of the cosmic microwave background by more than one order of magnitude. 
\end{abstract}

\maketitle

\section{Introduction}
The canonical cosmological model assumes that dark matter is comprised of a cold gas of weakly interacting particles. Despite its simplicity, this minimal scenario provides an excellent fit to both cosmic microwave background (CMB) and large-scale structure measurements~\cite{Adam:2015rua, Ade:2015xua, Aghanim:2015xee, Alam:2016hwk}. However, a precise understanding of the fundamental nature of dark matter is missing and remains at the forefront in the current list of unsolved problems in modern physics.

Although dark matter is usually interpreted in terms of a new elementary particle, other alternatives exist. Black holes (BHs) produced prior to big bang nucleosynthesis (BBN), i.e., primordial black holes (PBHs), represent such an alternative --- remarkably, this solution is as old as particle dark matter~\cite{Chapline:1975}. In particular, this scenario has recently attracted much attention~\cite{Bird:2016dcv, Clesse:2016vqa, Clesse:2016ajp, Sasaki:2016jop, Raidal:2017mfl, Ali-Haimoud:2017rtz, Kavanagh:2018ggo} in the context of the LIGO and VIRGO discoveries of several binary BH mergers~\cite{Abbott:2016blz, TheLIGOScientific:2016pea, Abbott:2017vtc, Abbott:2016nmj, Abbott:2017oio}.

The idea that PBHs could be formed by strong accretion during the radiation-dominated epoch was first introduced five decades ago~\cite{Zeldovich:1967}. It was later suggested that initial fluctuations in the early Universe could give rise to a large number of gravitationally collapsed objects with masses above the Planck mass~\cite{Hawking:1971ei}, that however, would not grow substantially by accretion~\cite{Carr:1974nx}.
Very light PBHs would not survive until the present epoch, though; the prediction that any BH should emit particles with a black body spectrum~\cite{Hawking:1974rv, Hawking:1974sw} implies that PBHs with masses below $10^{-18} M_\odot$ (with $M_\odot$ the Sun's mass) would have evaporated on cosmological times scales~\cite{Hawking:1974rv, Hawking:1974sw, Page:1976df}. While the evaporation of such light PBHs would have interesting observational consequences~\cite{Khlopov:2008qy, Carr:2009jm, Calmet:2014dea}, only heavier PBHs could constitute a non-negligible fraction of the dark matter content of the Universe~\cite{Chapline:1975} (see, e.g., Refs~\cite{Belotsky:2014kca, Carr:2016drx, Sasaki:2018dmp} for recent reviews). Alternative mechanisms for the formation of PBHs have also been proposed, such as the collapse of domain walls~\cite{Sato:1981bf, Maeda:1981gw, Berezin:1982ur} or cosmic strings~\cite{Hogan:1984zb, Hawking:1987bn, Polnarev:1988dh}, first-order phase transitions~\cite{Crawford:1982yz, Hawking:1982ga, Kodama:1982sf, Hall:1989hr, Moss:1994iq, Konoplich:1999qq, Jedamzik:1999am, Khlopov:2000js}, or fragmentation of scalar condensates~\cite{Cotner:2016cvr, Cotner:2017tir, Cotner:2018vug}. Today, the most studied mechanism of PBH formation is gravitational collapse~\cite{Carr:1975qj, Ivanov:1994pa, Yokoyama:1995ex, Shibata:1999zs, Young:2014ana} of the fluctuations predicted in inflationary scenarios~\cite{Carr:1993aq, Carr:1994ar, Ivanov:1994pa, Yokoyama:1995ex, GarciaBellido:1996qt, Taruya:1998cz, Green:2000he, Bassett:2000ha, Pi:2017gih}.

Following the recent observations of BH mergers by the LIGO and VIRGO experiments~\cite{Abbott:2016blz, TheLIGOScientific:2016pea, Abbott:2017vtc, Abbott:2016nmj, Abbott:2017oio}, PBHs of several solar masses or heavier have been extensively examined and constrained by a number of different means (see, e.g., Refs.~\cite{Mack:2006gz, Ricotti:2007au, Josan:2009qn, Carr:2009jm, Capela:2013yf, Clesse:2016vqa, Carr:2016drx, Green:2016xgy, Bellomo:2017zsr, Kuhnel:2017pwq, Carr:2017jsz, Sasaki:2018dmp} for compilations of constraints, also at other mass scales): gravitational lensing~\cite{Green:2017qoa, Garcia-Bellido:2017xvr, Zumalacarregui:2017qqd, Garcia-Bellido:2017imq}, dynamical constraints~\cite{Monroy-Rodriguez:2014ula, Brandt:2016aco, Green:2016xgy, Li:2016utv, Koushiappas:2017chw, Kavanagh:2018ggo}, radio and X-rays measurements~\cite{Gaggero:2016dpq, Inoue:2017csr, Hektor:2018rul, Manshanden:2018tze}, and spectral and anisotropy distortions of the CMB~\cite{Tada:2015noa, Young:2015kda, Chen:2016pud, Ali-Haimoud:2016mbv,  Horowitz:2016lib, Blum:2016cjs, Poulin:2017bwe, Bernal:2017vvn, Nakama:2017xvq, Deng:2018cxb}. If energy is injected into the intergalactic medium (IGM) via the accretion of matter onto PBHs, the ionization and thermal evolution of the Universe could be affected and thus, both the spectral shape~\cite{Carr:1993aq, Carr:1994ar, Chluba:2012we, Kohri:2014lza} and the anisotropies of the CMB~\cite{Carr:1981, Ricotti:2007au, Ricotti:2007au, Chen:2016pud, Ali-Haimoud:2016mbv, Blum:2016cjs, Bernal:2017vvn, Horowitz:2016lib, Poulin:2017bwe}. Namely, the work of Ref.~\cite{Poulin:2017bwe} excludes PBH with masses $\gtrsim 2 \, M_\odot$ as the sole component of dark matter, although uncertainties associated with the accretion make the determination of a robust upper limit difficult. Moreover, Poisson-noise fluctuations could become too large to be consistent with the power spectrum on scales of the Lyman-$\alpha$ forest, setting an upper limit on PBH masses near a few times $10^4 \, M_\odot$~\cite{Afshordi:2003zb}, which has been recently improved to $M_{\rm PBH}\lesssim 60 \, M_\odot$~\cite{Murgia:2019duy}.                                                       

The presence of PBHs could also change the process known as reionization, during which the first sources in the Universe emitted ultraviolet photons that ionized the neutral hydrogen. Collectively, observations from the CMB observations, Lyman-$\alpha$ emission in star-forming galaxies, and of the Gunn-Peterson optical depth from bright quasars indicate that the Universe should be fully ionized by $z \sim 6$. Nevertheless, only observations of the CMB are currently sensitive to the early part of reionization, and to the epoch during which the first stars formed (referred to as the cosmic dawn). The CMB, however, is only sensitive to a cumulative line-of-sight effect (observed via modifications to the integrated optical depth), and thus cannot provide detailed information about the state of the IGM during these epochs. The only proposed method for understanding the nature of the Universe on these time-scales is 21cm cosmology, which attempts to observe the hyperfine transition between the singlet and triplet levels of neutral hydrogen. To date, only one experiment working has claimed a putative signal; using a single dish radio telescope, the \texttt{EDGES} (Experiment to Detect the Reionization Step) collaboration~\cite{Bowman:2018yin} claimed detection of a large absorption dip in the globally averaged signal, appearing around $z \sim 17$, but whose amplitude cannot be explained within the standard $\Lambda$CDM scenario. Determining the validity of this signal, however, is notoriously challenging, as foregrounds at these frequencies are expected to be at least $\sim 10^4$ times larger than the signal of interest. Thus, the 21cm contribution can only be discriminated by systematically removing spectrally smooth foreground components from the observation. Next-generation radio interferometers such as \texttt{HERA} (Hydrogen Epoch of Reionization  Array)~\cite{Beardsley:2014bea} and \texttt{SKA} (Square  Kilometer  Array)~\cite{Mellema:2012ht} will also attempt to observe these epochs. A major advantage of using an interferometer in these observations is that one also has direct information on the scales over which foregrounds emanate, allowing for a more robust removal of foregrounds and providing additional spatial information on the 21cm signal. 

The field of 21cm cosmology should be considered particularly important for those in the field of particle physics, as the wealth of available information might allow physicists to probe any new process which, e.g., modifies structure formation, the energy in the IGM, star formation and evolution, and photon propagation. Studies have been performed in the context of various dark matter candidates, for example, illustrating unprecedented sensitivity to a wide array of dark matter candidates~\cite{Valdes:2012zv, Sitwell:2013fpa, Evoli:2014pva, Carucci:2015bra, Lopez-Honorez:2016sur, Berlin:2018sjs, DAmico:2018sxd, Liu:2018uzy, Mitridate:2018iag, Pospelov:2018kdh, AristizabalSierra:2018emu, Boddy:2018wzy, Kovetz:2018zan, Barkana:2018cct, Barkana:2018lgd, Fialkov:2018xre, Escudero:2018thh, Schneider:2018xba, Safarzadeh:2018hhg, Lidz:2018fqo, Lopez-Honorez:2018ipk}. Existing works looking into the effects of PBHs in 21cm cosmology have independently focused on understanding the effects arising from PBH accretion on the globally averaged signal~\cite{Hektor:2018qqw}, Poisson-noise on the  power spectrum~\cite{Gong:2017sie, Gong:2018sos}, and local modifications to the medium around PBHs on the globally averaged signal and the angular power spectrum~\cite{Tashiro:2012qe, Bernal:2017nec}. In this work, we compare and contrast these three effects in the context of $\gtrsim \mathcal{O}(M_\odot)$ PBHs, illustrating that the only process relevant for $M_{\rm PBH} \lesssim 10^3 \, M_\odot$ is the heating and ionization of the IGM arising from X-rays produced during accretion. Using the 21cm power spectrum, which offers a far more robust and powerful probe the globally averaged signature, we illustrate the near-future sensitivity of both HERA and SKA. 

The structure of the paper is as follows. In \Sec{sec:21cmcosmo} we review the basics of 21cm cosmology, in particular focusing on the computation of the 21cm brightness temperature and power spectrum. In \Sec{sec:pbh_acc} we discuss the impact that PBHs have on the observable 21cm signal; in particular we study global heating and ionization as a consequence of accretion around PBHs (\Sec{sec:globalPBH}), and how the process of accretion heats and ionizes the local medium surrounding the PBH (\Sec{sec:localPBH}), and we also address how Poisson-noise induced in the power spectrum by the discrete nature of PBHs modifies the signal arising in minihalos (\Sec{sec:shot_N}). We comment on the sensitivity analysis in \Sec{sec:methods}, and then present the results and conclude in \Sec{sec:conclusions}.

\section{21cm cosmology}
\label{sec:21cmcosmo}

Below, we present a brief overview of the fundamentals of 21cm cosmology. A more extensive overview can be found, e.g., in Refs.~\cite{Madau:1996cs, Furlanetto:2006jb, Pritchard:2011xb, Furlanetto:2015apc}.

The intensity of the redshifted 21cm line provides direct information about the fraction of neutral hydrogen residing in excited and ground states. This is conventionally characterized by the so-called spin temperature $T_S$, and is formally defined as
\begin{equation}
\frac{n_1}{n_0} = 3 \, e^{- T_0/ T_S} ~,
\end{equation}
where $n_0$ and $n_1$ are the number density of neutral hydrogen atoms in the ground (singlet) and excited (triplet) states, $k_B T_0 = h \nu_0$ is the energy of the 21cm photon, and the factor of 3 comes from the degeneracy of the triplet excited state. The spin temperature of the 21cm line is controlled by three processes: $\emph{(i)}$ absorption of and stimulated emission induced by a background radiation field, $\emph{(ii)}$ collisions of neutral hydrogen with hydrogen atoms, free protons, or free electrons, and $\emph{(iii)}$ indirect excitations/de-excitations induced by scattering with ambient Lyman-$\alpha$ photons. Consequently, the spin temperature can be expressed directly in terms of temperatures characterizing the efficiency of each of these processes, 
\begin{equation}\label{eq:spinT}
T_S = \frac{T_{\rm R} + y_k \, T_k + y_\alpha \, T_\alpha}{1 + y_k + y_\alpha} ~,
\end{equation}
where $T_{\rm R}$, $T_k$, and $T_\alpha$ are the temperature of the background radiation, kinetic temperature of the gas, and the temperature associated with the intensity of the Lyman-$\alpha$ emission (for all environments of interest, it is sufficient to take $T_\alpha \simeq T_k$), and $y_k$ and $y_\alpha$ are their respective couplings. 

The collisional coupling, $y_k$, is given by
\begin{equation}
\label{eq:yk}
y_k = \frac{T_0}{A_{10} \, T_k} \, n_{\rm H} \, \left( x_{\rm H} \, C_{\rm HH} + (1 - x_{\rm H}) \, C_{\rm eH} + (1 - x_{\rm H}) \, C_{\rm pH} \right) ~, 
\end{equation}
where $A_{10} = 2.85 \times 10^{-15} {\rm s}^{-1}$ is the Einstein coefficient of spontaneous emission rate, $n_{\rm H}$ is the hydrogen comoving number density, $x_{\rm H}$ is the neutral fraction of hydrogen, and $C_{\rm HH}$, $C_{\rm eH}$ and $C_{\rm pH}$ are the de-excitation rates due to collisions between a hydrogen atom and another hydrogen atom, an electron, or a proton, respectively. These quantities are computed using the fitting formulas of Ref.~\cite{Kuhlen:2005cm}. 

The Lyman-$\alpha$ coupling, $y_\alpha$, can be written in terms of the Lyman-$\alpha$ flux $J_{\alpha}$,
\begin{equation}
\label{eq:yalpha}
y_\alpha = \frac{16\pi^2 \, e^2 \, f_{12}}{27\, m_e \, c \, A_{10}} \, \frac{T_0}{T_k} \, J_{\alpha} ~,
\end{equation}
where $e$ and $m_e$ are the charge and mass of the electron, and $f_{12} = 0.416$ is the oscillator strength of the Lyman-$\alpha$ transition.

Rather than working directly with observed intensity, it is conventional for radio astronomers to use the effective brightness temperature, $T_b(\nu)$, which is proportional to the specific intensity (or spectral radiance) of a blackbody in the Rayleigh-Jeans limit, $\nu \ll T_b(\nu)$. The change in intensity relative to the background radiation field is thus given by the differential brightness temperature:
\begin{equation}
\delta T_b(\nu) = \frac{T_S - T_{\rm R}}{1 + z} (1 - e^{-\tau_{\nu_0}}) ~,
\label{eq:Tb}
\end{equation}
where $\tau_{\nu_0}$ is the optical depth of the 21cm line. Throughout this work we will implicitly assume the background radiation field to simply be the CMB, $T_{\rm R} = T_{\rm CMB}$. We note, however, that in the light of the abnormally deep absorption trough observed by EDGES~\cite{Bowman:2018yin}, several scenarios predicting an unexpected excess of radio background have been explored~\cite{Feng:2018rje, Ewall-Wice:2018bzf, Fraser:2018acy, Mirocha:2018cih, Pospelov:2018kdh, Dowell:2018mdb, Jana:2018gqk, Fialkov:2019vnb, Ewall-Wice:2019may} (see, however, Ref.~\cite{Sharma:2018agu}).

\section{The effects of primordial black holes}
\label{sec:pbh_acc}

The existence of a population of $M_{\rm PBH} \gtrsim M_\odot$ PBHs has the potential to modify the 21cm signal in a number of different ways. In order to understand exactly how and where these effects would arise, let us consider for the moment a single isolated PBH. Matter surrounding the PBH is accreted at some rate $\dot{M}_{\rm PBH}$. The in-falling matter heats and ionizes in a localized region surrounding the PBH. Some of the energy gained via accretion is radiated by the newly formed plasma, primarily in the form X-rays produced by thermal synchrotron and bremsstrahlung emission, and modified by comptonization. Additionally, energy resulting from the in-fall of matter can be viscously dissipated, directly heating the accreting matter rather than radiating X-rays. This type of energy dissipation is at the heart of so-called advection-dominated accretion flow (ADAF) (see, e.g., Ref.~\cite{Yuan:2014gma} for a review). It has recently been argued~\cite{Poulin:2017bwe} that accretion onto $\sim\mathcal{O}(M_\odot)$ PBHs at high redshifts is likely to proceed via a thick inflated disk, best described within the ADAF framework --- we will thus follow Ref.~\cite{Poulin:2017bwe} in adopting this formalism. Typically, ADAF accreting models radiate extremely inefficiently (i.e., only a small fraction of the accreted energy is radiated and escapes into the IGM), while the remainder is dissipated locally in the form of heat through viscous interactions of the plasma. The relative efficiency with which X-rays are emitted in the IGM can be characterized by a parameter $\epsilon$ ($\epsilon \rightarrow 1$ being the limit that the accretion medium radiates perfectly, and $\epsilon \rightarrow 0$ being the limit in which most of the accreted energy goes into local heating via advection), which depends intimately on the geometry and strength of accretion. X-rays that escape into the medium  have a large mean free path and would deposit their energy approximately uniformly on all scales, affecting both the temperature and ionization fraction of the IGM. On the other hand, the remainder of the energy is deposited locally (and is eventually swallowed by the PBH), modifying the temperature and ionization profile of the medium directly surrounding the PBH. Thus, accretion onto a single BH modifies the state of the IGM both globally (\Sec{sec:globalPBH}) and locally (\Sec{sec:localPBH}), effects which for the sake of simplicity we treat separately in what follows. 

In addition to the effects arising from accretion, the presence of a large population of PBHs can modify structure formation. The discrete nature of massive PBHs implies that shot noise fluctuations in the PBH number density are not necessarily negligible. These fluctuations manifest as isocurvature perturbations in the linear matter power spectrum, leading to an enhancement in structure at small scales~\cite{Afshordi:2003zb}. For viable populations of PBHs (i.e., those evading current constraints), the effect is only expected to be pronounced for halos below the star-forming limit --- these halos are typically referred to as minihalos. Within the context of $\Lambda$CDM, minihalos contribute minimally to the 21cm signal \cite{Iliev:2002gj} (the signal is expected to be at the level of $\mathcal{O}(1) \,$mK in $\Lambda$CDM). However, it has been argued that the enhancement in the number density of these halos by a population of PBHs can make their contribution observable~\cite{Gong:2017sie, Gong:2018sos}. We address this claim in \Sec{sec:shot_N}, and show that the contribution of minihalos is typically suppressed when the global effects of accretion are self-consistently included in the calculation.

\subsection{Global heating and ionization}
\label{sec:globalPBH}

Here we address the global impact of accretion by PBHs on the IGM, i.e., the effect that arises from radiation that escapes the local vicinity of PBHs and deposits its energy approximately uniformly in the IGM. In what follows we consider the simplifying assumption of a monochromatic mass function for PBHs, broadly used in the literature. Note, however, that this is not necessarily a realistic assumption, since many mechanisms of PBH formation in the early Universe would generically lead to extended mass distributions, which could bias the obtained constraints~\cite{Carr:2016drx, Green:2016xgy, Carr:2017jsz, Bellomo:2017zsr}.
 
The rate of energy injected into the medium per unit volume is given by \cite{Ali-Haimoud:2016mbv, Poulin:2017bwe}
\begin{equation}\label{eq:enginj}
\left(\frac{dE}{dV\,dt}\right)_{\rm inj} = L_{\rm acc} \, n_{\rm PBH} = L_{\rm acc} \, \frac{f_{\rm PBH} \, \rho_{\rm DM}}{M_{\rm PBH}} ~,
\end{equation}
where $L_{\rm acc}$ is the luminosity of radiation emitted from matter accreting onto a PBH of mass $M_{\rm PBH}$, $n_{\rm PBH}$ is the number density of PBHs of mass $M_{\rm PBH}$, $\rho_{\rm DM}$ is the density of dark matter, and $f_{\rm PBH}$ is the fraction of dark matter that is comprised of PBHs of mass $M_{\rm PBH}$. The accretion luminosity can be expressed in terms of the product of the accretion rate $\dot{M}_{\rm PBH}$ and the radiative efficiency $\epsilon$, 
\begin{equation}\label{eq:lacc}
L_{\rm acc} = \epsilon \, \dot{M}_{\rm PBH} ~.
\end{equation}
In general, both the accretion rate and the radiative efficiency are complicated unknown functions of both the PBH properties and the evolution of the medium. However, various approximations have been developed that can be used to estimate each of these functions.

We begin by addressing the radiative efficiency factor $\epsilon$ in \Eq{eq:lacc}, which describes the fraction of energy radiated away into the IGM. In principle, it depends on the accretion rate, accretion geometry, and the properties of the medium. Here, we assume that the radiative cooling is inefficient and the dynamics of the accreting gas are controlled by advective currents (ADAF)~\cite{Ichimaru:1977uf, Rees:1982pe, Narayan:1994xi, Blandford:1998qn, Narayan:2002ss, Pellegrini:2005pi, Xie:2012rs, Yuan:2014gma}. It is worth noting that if the radiative cooling is efficient and forms a thin accretion disk (rather than the thick disk or tori that may appear in the ADAF scenario), the radiative efficiency could be orders of magnitude larger~\cite{Shakura:1972te}. For the efficiency function, we adopt the fitting formula obtained in Ref.~\cite{Xie:2012rs},
\begin{equation}
\label{eq:adaffit}
\epsilon = \epsilon_0 \, \left(\frac{100 \, \dot{M}_{\rm PBH}}{\dot{M}_{\rm Edd}} \right)^a ~,
\end{equation}
where $\dot{M}_{\rm Edd}= 10 \, L_{\rm Edd} = 10 \times 4 \pi \, G M_{\rm PBH} \, m_p /\sigma_{\rm T}\simeq 1.28 \times 10^{39} (M_{\rm PBH}/M_\odot)$ ergs/s is the Eddington rate (and $L_{\rm Edd}$ is the Eddington luminosity), and the functional form and coefficients are indicated in \Tab{table:adaf}, assuming the viscous parameter to be $\alpha = 0.1$ and the parameter which dictates the fraction of dissipation that heats the electrons directly to be $\delta = 0.1$. Larger (smaller) values of $\delta$ are expected to increase (decrease) the radiative efficiency by a factor of a few in the most extreme scenarios.

\begin{table}
	\setlength\extrarowheight{5pt}
	\begin{center}
		\begin{tabular*}{0.4\textwidth}{c @{\extracolsep{\fill}} c c}
			\hline
			$\dot{M}_{\rm PBH} / \dot{M}_{\rm Edd}$ & $\epsilon_0$ & $a$ \\ 
			\hline\hline
			$\lesssim 9.4 \times 10^{-5}$ & 0.12 & 0.59 \\
			\hline
			$9.4\times 10^{-5} - 5 \times 10^{-3}$ & 0.026 & 0.27 \\
			\hline
			$5\times 10^{-3} - 6.6 \times 10^{-3}$ & 0.50 & 4.53\\
			\hline
		\end{tabular*}
	\end{center}
	\caption{Fit coefficients characterizing the radiative efficiency of ADAF accretion, described by \Eq{eq:adaffit}, in terms of the net accretion rate. From Ref.~\cite{Xie:2012rs}, for $\delta = 0.1$ and for the viscous parameter $\alpha = 0.1$.}
	\label{table:adaf}
\end{table}

We now shift our attention to the accretion rate. Accretion onto a point mass moving through some homogeneous medium was first studied by Hoyle and Lyttleton in the first half of the $20^{\rm th}$ century~\cite{hoyle1939effect, hoyle1940accretion, hoyle1940physical, hoyle1941accretion}. Nevertheless, this early treatment only considered the gravitational effects (neglecting, for example, the thermodynamic pressure), and was thus only useful as a first-step approximation within a particular limiting regime (far away from the point source and with a reasonably large relative velocity). Bondi and Hoyle were later able to correct for the previously neglected effect of pressure and calculated an exact solution to the problem of spherical accretion onto a stationary point mass in a homogenous medium~\cite{bondi1944mechanism}, 
\begin{equation}
\label{eq:bondirate}
\dot{M}_{\rm Bondi-Hoyle} = 4 \pi \, \lambda \, \rho_\infty \, \frac{(G M_{\rm PBH})^2}{c_{s,\infty}^3} ~,
\end{equation}
where $\rho_\infty$ and $c_{s,\infty}$ are the density and speed of sound far away from the point source, $G$ is the gravitational constant, and $\lambda$ is an ${\cal O}(1)$ parameter quantifying non-gravitational effects~\cite{bondi1944mechanism}, which also allows to treat non-spherical accretion. Given that simulations suggest that radiative outflows reduce the amount of matter that ultimately accretes in the inner region, potentially reducing the observed luminosity by a factor of $\mathcal{O}(100)$~\cite{Xie:2012rs}, we account for this by taking $\lambda$ to be $0.01$.

In order to address the problem of accretion onto a BH moving with respect to the ambient medium with relative velocity $v_{\rm rel}$, Bondi suggested the substitution of $c_{s,\infty}$ by an effective velocity $\sqrt{c_{s,\infty}^2 + v_{\rm rel}^2}$~\cite{Bondi:1952ni}. The relative velocity between PBHs and the ambient medium can be described by the relative velocity between dark matter and baryons. Unfortunately, deriving the relative dark matter-baryon velocity at scales relevant for accretion is non-trivial~\cite{Tseliakhovich:2010bj, Dvorkin:2013cea}. On linear scales, the velocity field is Gaussian and thus, the modulus of the three-velocity field follows a Maxwellian distribution, with the square root of the variance approximated as~\cite{Ali-Haimoud:2016mbv}
\begin{equation}
\sqrt{\left<v_L^2\right>} \simeq {\rm min}\left[1, \, \frac{1+z}{1000} \right] \times 30 \, \rm{km/s} ~.
\end{equation}
The accretion luminosity is the result of averaging over the distribution of the relative velocities of PBHs with respect to the ambient gas. The accretion rate is inversely proportional to $\left(c_{s,\infty}^2 + v_{\rm rel}^2\right)^{3/2}$ (after the substitution above)~\cite{Ricotti:2007au, Ali-Haimoud:2016mbv} and, for the usual (simplified) case considered in the literature, the individual PBH accretion luminosity is proportional to the square of the accretion rate (i.e., $L_{\rm acc} \propto \dot{M}_{\rm PBH}^2$). It is important to emphasize that this applies only to scenarios in which $\epsilon \propto \dot{M}_{\rm PBH}$. Generalizing the accretion rate to scenarios in which the radiative efficiency has an arbitrary power law dependence (i.e., $\epsilon \propto \dot{M}_{\rm PBH}^a$) is obtained by expressing the effective velocity as 
%
\begin{equation} 
	\label{eq:veffparam}
	v_{\rm eff} \equiv \left< \frac{1}{\left(c_{s,\infty}^2 + v_{\rm rel}^2\right)^{3 \, (1+a)/2}}\right>^{-\frac{1}{3 \, (1+a)}}  = c_{s,\infty}  \,  \left[ \left(\frac{3}{2}\right)^{\frac{3}{2}} \, U\left(\frac{3}{2}, 1 - \frac{3 \, a}{2}, \frac{3}{2} \mathcal{M}^{-2}\right) \, \mathcal{M}^{-3} \right]^{- \frac{1}{3 \, (1+a)}}
\end{equation}
with $\mathcal{M} \equiv \frac{\sqrt{\left<v_L^2\right>}}{c_{s,\infty}}$ and $U (x ; y ; z)$ the confluent hypergeometric function of second kind, or Tricomi's function. Up to first order, the two limiting cases are:
\begin{equation}
v_{\rm eff} \simeq
\begin{cases}
 c_{s, \infty} \, {\cal{M}}^{1/(1+a)} \, \left[3 \, \sqrt{\frac{3}{2 \, \pi}} \, B\left(\frac{3 \, a}{2} , \frac{3}{2}\right) \right]^{-\frac{1}{3 \, (1+a)}}  & , \hspace{5mm} {\cal{M}} \gg 1 \\
 c_{s, \infty} & , \hspace{5mm} {\cal{M}}  \ll  1  ~,
\end{cases} 
\end{equation}
with $B(x, y)$ the beta function. Notice that this expression reproduces the limiting case in which $a = 1$ found in Refs.~\cite{Ricotti:2007au, Ali-Haimoud:2016mbv}.

Therefore, using the effective velocity rather than the speed of sound, the accretion rate can be written as
\begin{equation}
\label{eq:BHrate}
\dot{M}_{\rm PBH} = 4 \pi \, \lambda \, \rho_\infty \, \frac{(G M_{\rm PBH})^2}{v_{\rm eff}^3} ~.
\end{equation}
Even if more accurate expressions have been obtained~\cite{Mellah:2015sja}, this modified accretion rate provides a reasonable description of the problem. It is worth noting that the Bondi accretion rate is strictly valid only for spherical accretion. It was recently argued, however, that accretion onto PBHs with $M_{\rm PBH} \gtrsim M_\odot$ likely results in the formation of an accretion disk~\cite{Poulin:2017bwe}. Note that this statement ran contrary to previous assumptions that PBH accretion was spherical~\cite{Ricotti:2007au, Horowitz:2016lib, Ali-Haimoud:2016mbv, Blum:2016cjs}. Given that there exists a relatively large amount of uncertainty in the dynamics of disk accretion, some of which serves to increase and some of which serves to decrease the relative accretion rate~\cite{Krumholz:2004vj, Krumholz:2005pb, Mellah:2015sja}, \Eq{eq:BHrate} can likely be thought of as a reasonable ball-park estimate. Therefore, despite all these caveats regarding the potential legitimacy of Eqs.~(\ref{eq:veffparam}) and~(\ref{eq:BHrate}), though, we will follow previous works~\cite{Ali-Haimoud:2016mbv, Poulin:2017bwe} in adopting the expression of $v_{\rm eff}$ for $a=1$. 

Finally, it is important to note that the quantity of interest is not the rate of energy injected into the medium,  given in \Eq{eq:enginj}, but rather the rate of energy deposited into the medium. These two quantities are related by functions known as energy deposition factors, 
\begin{equation}
\label{eq:eninj}
\left(\frac{dE_c}{dV dt}\right)_{{\rm dep}} = f_c(z) \, \left(\frac{dE}{dV dt}\right)_{\rm inj} \equiv n_b \, \epsilon_c^{\rm PBH} ~,
\end{equation}
where $\epsilon_c^{\rm PBH}$ represents the energy deposition rate per baryon, and the subscript $c$ denotes the channel in which energy is deposited --- i.e., ionization of hydrogen (HI) or helium (HeI), heating of the medium (heat), or excitations (${\rm Ly}\alpha$). These energy deposition factors are computed by using the following expression:
\begin{equation}\label{eq:energydeposit}
f_c(z) = \frac{H(z)\int \frac{d\ln(1+z^\prime)}{H(z^\prime)} \int d\omega \, T_c(z,z^\prime,\omega) \, L_{\rm acc}(z^\prime, \omega)}{\int d\omega \, L_{\rm acc}(z, \omega)} ~,
\end{equation}
where the $T_c(z,z^\prime,\omega)$ are the transfer functions taken from Ref.~\cite{Slatyer:2015kla}. As is done in Ref.~\cite{Poulin:2017bwe}, the spectrum of the luminosity is taken from ADAF models of Ref.~\cite{Yuan:2014gma}; namely, we adopt a simple parameterization given by 
\begin{equation}\label{eq:L_spec}
L(\omega) \propto \Theta(\omega - \omega_{\rm min}) \, \omega^{\beta} \, \exp(-\omega / \omega_s) ~,
\end{equation}
where $\omega_{\rm min} \equiv (10 \, M_\odot / M_{\rm PBH})^{1/2}$ eV, $\omega_s \sim \mathcal{O}(m_e)$ (taken explicitly to be $\omega_s = 200$~keV), and $\beta = -1$ (with reasonable values of $\beta \in [-1.3, -0.7]$). In principle, the exact form of the spectrum changes with the accretion rate. However, the results found here are largely independent of these details. 

Eqs.~(\ref{eq:lacc}) and~(\ref{eq:L_spec}) provide the ingredients necessary to compute the energy injected into the IGM from a population of PBHs as a function of redshift, and \Eq{eq:energydeposit} describes how and where the energy is deposited. With this in hand, we now turn to incorporating this energy injection into the equations governing the evolution of the ionization fraction, the gas temperature, and the Lyman-$\alpha$ flux.

\begin{figure}
	\centering
	\includegraphics[width=0.8\textwidth]{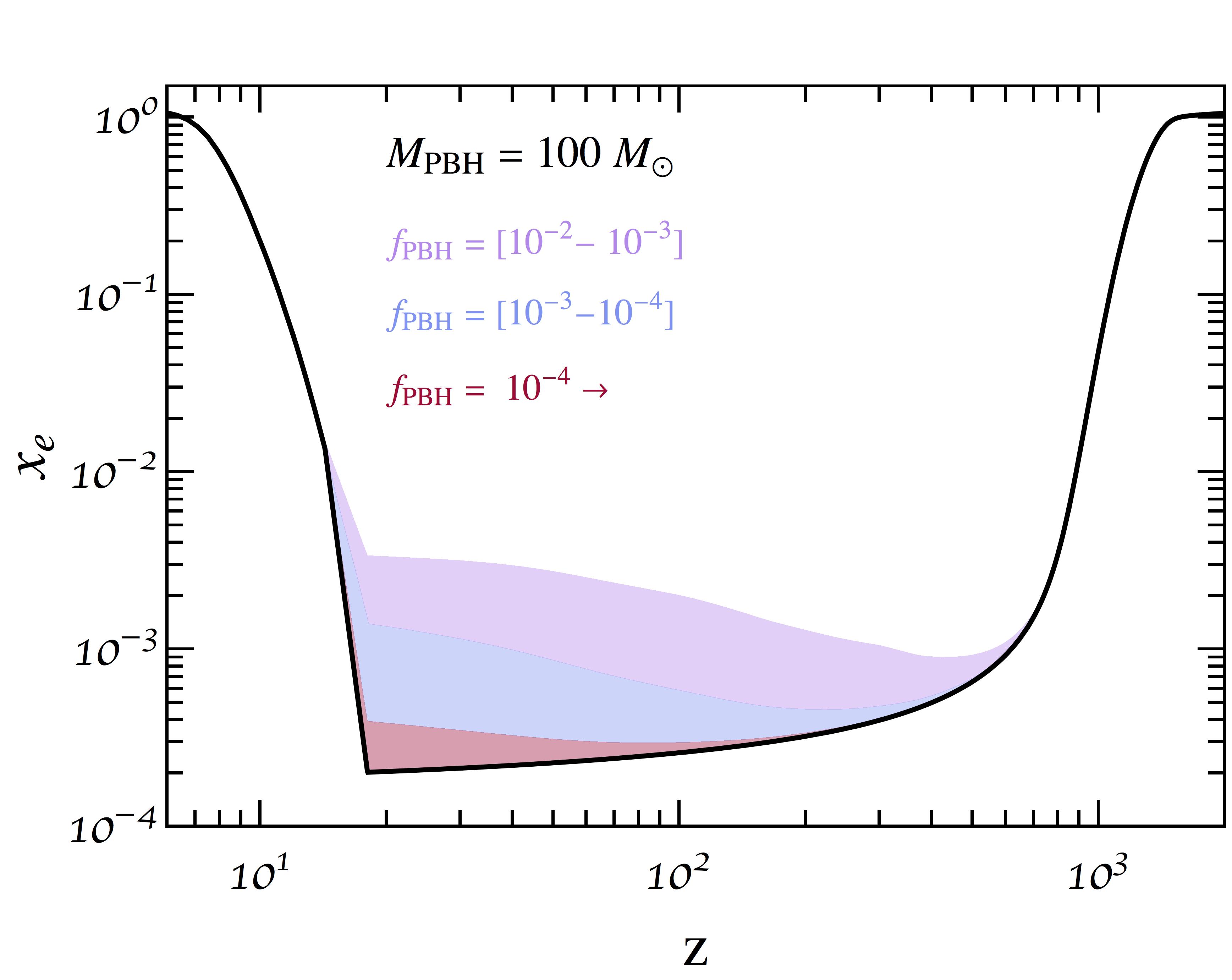}
	\caption{Free electron fraction, $x_e$, as a function of redshift, including the contribution of a monochromatic PBH population with mass $M_{\rm PBH} = 100 \, M_\odot$, for different PBH dark matter fractions $f_{\rm PBH} = \left( 10^{-2},  10^{-3},  10^{-4}, \lesssim 10^{-4} \right)$. The standard scenario with  $f_{\rm PBH}=0$ is  denoted by the solid black line. We use fiducial astrophysical parameters: $\left(\zeta_{\rm UV}, \, \zeta_{\rm X}, \, T_{\rm min}, \, N_\alpha\right) =  \left(50, \, 2 \times 10^{56} \, M_\odot^{-1}, \, 5 \times 10^{4} \, K, \, 4000\right)$; see \Sec{subsec:numer}.}
	\label{fig:xe} 
\end{figure}

\begin{figure}
	\centering
	\includegraphics[width=0.8\textwidth]{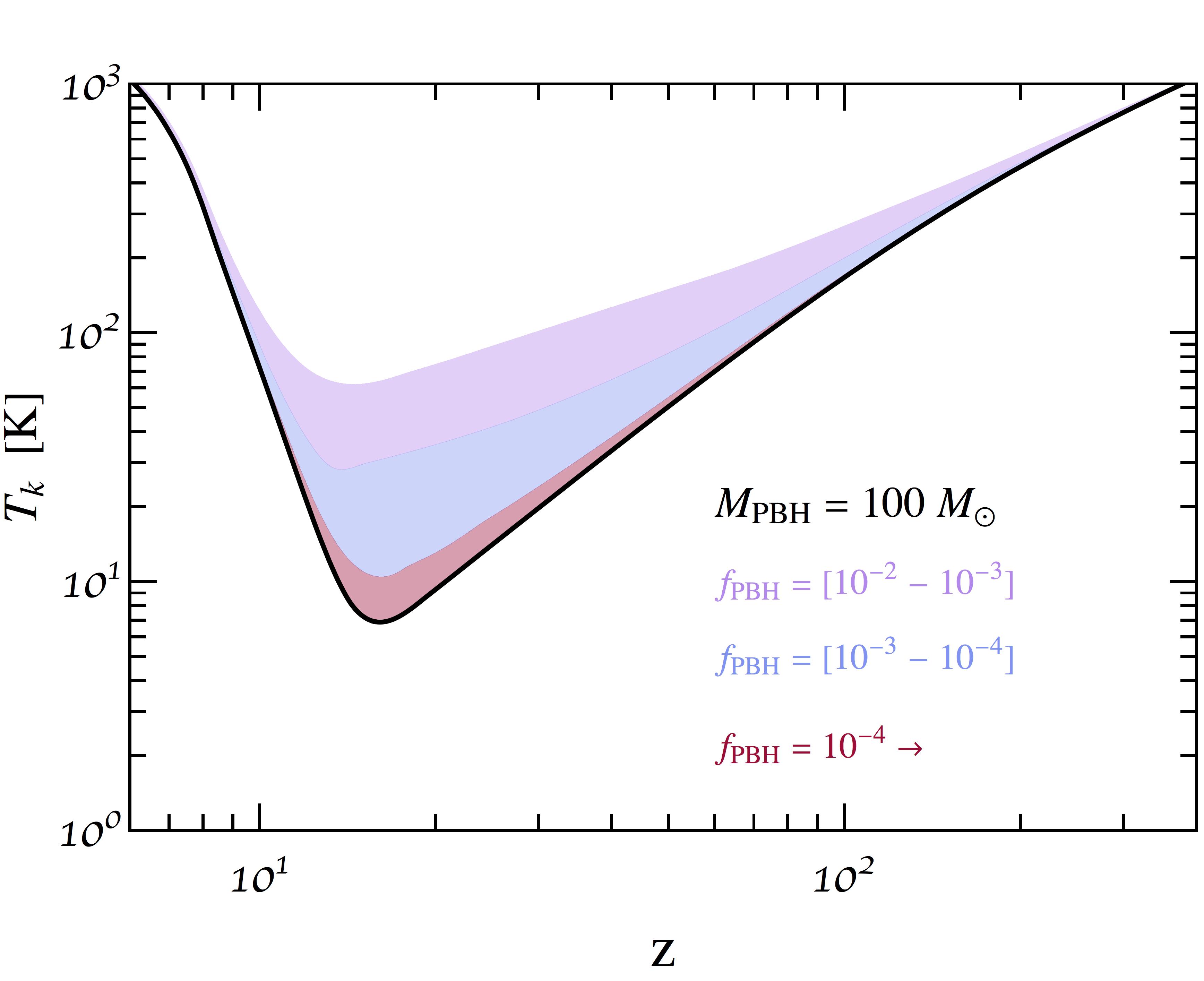}
	\caption{Kinetic temperature of the gass, $T_k$, as a function of redshift, including the contribution of a monochromatic PBH population with mass $M_{\rm PBH} = 100 \, M_\odot$, for different PBH dark matter fractions $f_{\rm PBH} = \left( 10^{-2},  10^{-3},  10^{-4}, \lesssim 10^{-4} \right)$. The standard scenario with  $f_{\rm PBH}=0$ is denoted by the solid black line. We use fiducial astrophysical parameters: $\left(\zeta_{\rm UV}, \, \zeta_{\rm X}, \, T_{\rm min}, \, N_\alpha\right) =  \left(50, \, 2 \times 10^{56} \, M_\odot^{-1}, \, 5 \times 10^{4} \, K, \, 4000\right)$; see \Sec{subsec:numer}.}
	\label{fig:Tk}
\end{figure}

The quantity $x_e({\bf x}, z)$, governing the evolution of the local ionized fraction of the neutral IGM, is given by
\begin{equation}
\label{eq:xe}
\frac{dx_e}{dz} = \frac{dt}{dz} \left(\Lambda_{\rm ion} - \alpha_{\rm A} \, C \, x_e^2 \, n_b \, \mathfrak{f}_{\rm H} \right) ~,
\end{equation}
where
$n_b = \bar{n}_{b, 0} (1+z)^3 (1 + \bar{\delta_b}({\bf x}, z))$ is the baryon number density, $\Lambda_{\rm ion}$ is the ionization rate, $\alpha_{\rm A}$ is the case-A recombination coefficient, $\mathfrak{f}_{\rm H} = n_{\rm H}/n_b$ is the hydrogen number fraction and $C\equiv \langle n_e^2 \rangle / \langle n_e \rangle^2$ is the clumping factor (set to one as default), with $n_e$ the electron number density. Ionization from the PBH accretion mechanism would lead to an additional contribution of the form
\begin{equation}
\Lambda_{\rm ion}^{\rm PBH} =\mathfrak{f}_{\rm H} \, \frac{\epsilon_{\rm HI}^{\rm PBH}}{E_{\rm HI}} + \, \mathfrak{f}_{\rm He} \, \frac{\epsilon_{\rm HeI}^{\rm PBH}}{E_{\rm HeI}} ~,
\end{equation}
where $\mathfrak{f}_{\rm H} = n_{\rm{H}}/n_{\rm{b}}$ and $\mathfrak{f}_{\rm He} = n_{\rm{He}}/n_{\rm{b}}$ are the hydrogen and helium number fractions, and $E_{\rm HI, HeI}$ are the ionization energies for hydrogen and helium. Here, we neglect the effect of secondary ionizations\footnote{Since the differential brightness temperature is proportional to the neutral hydrogen fraction, rather than the free electron fraction, the effect of ionizations from PBH accretion is expected to be significantly subdominant to that of heating. Thus, neglecting secondary ionizations is a safe assumption.}. The evolution of the kinetic temperature of the gas $T_k$ is computed via
\begin{equation}
\label{eq:dTkdzacc}
\frac{dT_k}{dz} = \frac{2 \, T_k}{3 \, n_b} \frac{dn_b}{dz} - \frac{T_k}{1 + x_e} \frac{dx_e}{dz} + \frac{2}{3 \, k_B \, (1 + x_e)} \frac{dt}{dz} \sum_j \epsilon_{{\rm heat}, j} ~,
\end{equation}
where the last term accounts for the heating/cooling processes, with $\epsilon_{{\rm heat}, j}$ the heating rate per baryon for the process $j$ (Compton cooling, X-ray heating and PBH heating). 

Finally, there is also a contribution to the Lyman-$\alpha$ flux resulting from collisional excitations due to energy deposition by PBH accretion in the IGM, which can be written as
 \begin{equation}
J_{\alpha,\mathrm{PBH}}=\frac{c \, n_b}{4\pi} \,\frac{\epsilon_{{\rm Ly}\alpha}^{\rm PBH}}{h\nu_{\alpha}} \, \frac{1}{H(z) \, \nu_\alpha} ~,
\label{eq:Jalph}
\end{equation}
where $\nu_{\alpha}$ is the emission frequency of a Lyman-$\alpha$ photon.
 
The aforementioned modifications to the free electron fraction, temperature of the gas, and Lyman-$\alpha$ flux are incorporated by modifying the publicly available codes \texttt{cosmorec/Recfast++}~\cite{Chluba:2010ca, AliHaimoud:2010ab, Chluba:2010fy}, relevant at high redshifts when astrophysical mechanisms can be neglected, and {\tt 21cmFAST}~\cite{Mesinger:2010ne}, relevant at low redshifts when astrophysical mechansims, such as ionization and heating from stars and X-ray binaries, are relevant. Figures~\ref{fig:xe} and \ref{fig:Tk} show the redshift evolution of the free electron fraction and the kinetic gas temperature for a population of PBHs with mass $M_{\rm PBH} = 100 \, M_{\odot}$ and different relative abundances $f_{\rm PBH}$.\footnote{This range of values is simply intended to illustrate the dependence of these observables on $f_{\rm PBH}$. Note that the cases with the largest abundances are in tension with CMB constraints~\cite{Poulin:2017bwe}.} Notice that the effect of PBHs accretion on the free electron fraction in \Fig{fig:xe} is clearly visible: the presence  of the extra heating and ionization terms from PBHs accretion changes the redshift evolution of $x_{\rm e}$, increasing this quantity from the early recombination era, below $z \sim 1000$, until the late reionization era. The kinetic gas temperature would also be increased by the presence of the energy injection in the IGM (see \Fig{fig:Tk}). Similar to the case in which there is energy injection from dark matter annihilations~\cite{Evoli:2014pva, Lopez-Honorez:2016sur}, PBH accretion leads to an earlier and more uniform heating of the IGM, which is larger for an increasing fraction of dark matter in the form of PBHs, until stellar sources turn on and start to ionize the medium (around $z \sim 15$ in \Fig{fig:xe}).\footnote{Note that although the spatial and redshift PBH distribution follows that of matter, it is different from the distribution of X-ray sources, i.e., star-forming halos beyond a threshold for atomic cooling.} These results illustrate that even small abundances of PBHs could have dramatic effects on the properties of the IGM.

Before continuing, we would like to emphasize that the treatment of accretion adopted in this work is rather conservative. For the redshifts relevant for 21cm cosmology, the conditions necessary for the formation of accretion disks around PBHs seem likely. Within the context of disk accreting models, ADAF accretion is among the lowest in the radiative efficiency of X-rays. Adopting a larger radiative luminosity or accretion rate would correspondingly enhance the observable signatures associated with global heating and ionization of the IGM.

\subsubsection{Global 21cm line signal}

Since the optical depth at the frequencies of interest is small, one can expand the exponential term in \Eq{eq:Tb} and express the global differential brightness temperature as
\begin{equation}
\delta T_b(\nu) \simeq 27 \, x_\textrm{H} \, (1 + \delta_b) \left( 1 - \frac{T_\textrm{CMB}}{T_S}\right) \left( \frac{1}{1+H^{-1} \partial v_r / \partial r} \right) \, \left( \frac{1+z}{10}\right)^{1/2} \left(\frac{0.15}{\Omega_{\rm m} h^2} \right)^{1/2} \left( \frac{\Omega_{\rm b} h^2}{0.023}\right)\,\textrm{mK} ~,
\label{eq:Tbdev}
\end{equation}
where $\delta_b$ is the baryonic overdensity, $H$ is the Hubble parameter, $\partial v_r / \partial r$ is the peculiar velocity gradient along the line of sight, which introduces redshift space distortions, and $\Omega_{\rm m}$ and $\Omega_{\rm b}$ are the matter and baryon abundances of the Universe.  A number of current and future experiments are devoted to detect the 21cm global signal averaged over all directions in the sky. Examples are \texttt{EDGES}  and  the future \texttt{LEDA} (Large Aperture Experiment to Detect the Dark Ages)~\cite{Greenhill:2012mn} and  \texttt{DARE} (Moon space observatory Dark Ages Radio Experiment)~\cite{Burns:2011wf}. The observation of an absorption profile located at a redshift of $z \sim 17$ has been recently claimed by the \texttt{EDGES} experiment, with an amplitude which is twice the maximum predicted within the context of the $\Lambda$CDM model~\cite{Bowman:2018yin}. This has motivated a larger number of studies in the literature~\cite{Munoz:2018pzp, McGaugh:2018ysb, Barkana:2018lgd, Barkana:2018cct, Fraser:2018acy, Kang:2018qhi, Yang:2018gjd, Pospelov:2018kdh, Costa:2018aoy, Slatyer:2018aqg, Falkowski:2018qdj, Munoz:2018jwq, Fialkov:2018xre, Berlin:2018sjs, DAmico:2018sxd, Safarzadeh:2018hhg, Hill:2018lfx, Clark:2018ghm, Cheung:2018vww, Hektor:2018qqw, Liu:2018uzy, Hirano:2018alc, Mitridate:2018iag, Mahdawi:2018euy, Feng:2018rje, Ewall-Wice:2018bzf, Mirocha:2018cih, Pospelov:2018kdh, Witte:2018itc, Dowell:2018mdb, Lopez-Honorez:2018ipk, Jana:2018gqk, Fialkov:2019vnb, Ewall-Wice:2019may}. 

\begin{figure}
	\centering
	\includegraphics[width=0.8\textwidth]{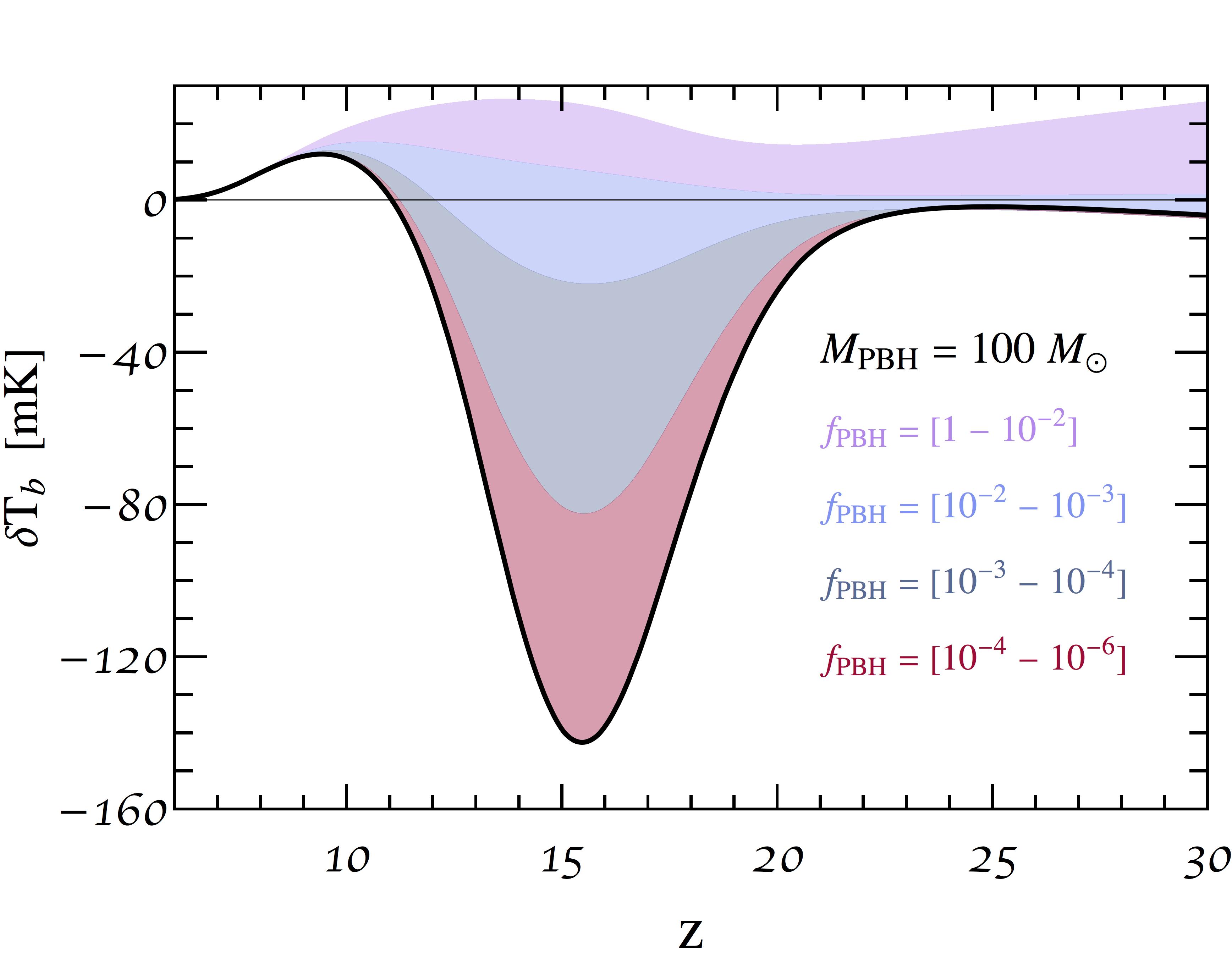}
	\caption{\label{fig:fig4} Global 21cm differential brightness temperature for various values of $f_{\rm PBH}$, assuming $M_{\rm PBH} = 100 \, M_\odot$ and different ranges for the PBH dark matter fraction. The standard scenario with $f_{\rm PBH} = 0$ is denoted by the solid black lines. We use fiducial astrophysical parameters: $\left(\zeta_{\rm UV}, \, \zeta_{\rm X}, \, T_{\rm min}, \, N_\alpha\right) =  \left(50, \, 2 \times 10^{56} \, M_\odot^{-1}, \, 5 \times 10^{4} \, K, \, 4000\right)$; see \Sec{subsec:numer}.}
\end{figure}

To compute the redshift evolution of the 21cm global signal we use of the publicly available code {\tt 21cmFAST}~\cite{Mesinger:2010ne}. Nevertheless, the code only evolves the IGM for redshifts $z \lesssim 35$. Thus, initial conditions for the mean ionization fraction $x_e$ and for the gas temperature $T_k$ are required inputs. We produce these inputs using a modified version of \texttt{cosmorec/Recfast++}~\cite{Chluba:2010ca, AliHaimoud:2010ab, Chluba:2010fy}.\footnote{Notice that these initial values are different from the ones in the default tables, as the presence of PBHs would change significantly the recombination process and its outputs.} The cumulative effect of a population of $100 \, M_\odot$ PBHs on the global differential 21cm brightness temperature is shown in \Fig{fig:fig4} for various PBH abundances. As discussed above, accretion onto PBHs significant increases both the free electron fraction and the temperature of the IGM (as shown in Figs.~\ref{fig:xe} and \ref{fig:Tk}). Consequently, the  effect of PBHs on the 21cm global differential brightness temperature can be summarized as a suppression of the absorption trough before X-ray heating dominates the signal. Notice the difference with the standard scenario, $f_{\rm PBH} = 0$, depicted by the solid black lines in \Fig{fig:fig4}.

It is also possible to measure the power spectrum of the differential brightness temperature, i.e., to extract the fluctuations in the all-sky averaged differential brightness temperature. This method could in principle provide a cleaner foreground removal. Indeed, this is the major goal of experiments such as \texttt{GMRT} (Giant Metrewave Radio Telescope), \texttt{LOFAR} (LOw Frequency ARray), \texttt{MWA} (Murchison Widefield Array) and \texttt{PAPER} (Precision Array for Probing the Epoch of Reionization), which has improved the upper limits at $z = 8.4$~\cite{Ali:2015uua}. As previously stated, future high-redshift 21cm experiments include \texttt{SKA} and \texttt{HERA}.\footnote{The 21cm line will also be observed at lower redshifts, predominately in the post-reionization era ($z\lesssim 3$), via the so-called intensity mapping technique, which will attempt to shed light on the spatial distribution emanating from dense clumps of neutral hydrogen~\cite{Wyithe:2007rq, Chang:2007xk, Loeb:2008hg, Villaescusa-Navarro:2014cma}. This is the target of the \texttt{GBT-HIM} project~\cite{Chang:2016npo}, \texttt{CHIME} (Canadian Hydrogen Intensity Mapping Experiment)~\cite{CHIME}, the \texttt{Tianlai} project~\cite{Chen:2015oga} and \texttt{SKA-mid} frequency (see, e.g., Ref.~\cite{Bull:2014rha}).} The dimensionless 21cm power spectrum is given by
\begin{equation}
\langle  \widetilde{\delta}_{21} (\mathbf{k}, z)  \widetilde{ \delta}_{21}^* (\mathbf{k}^\prime, z) \rangle \equiv (2\pi)^3 \delta^D (\mathbf{k} - \mathbf{k}^\prime) P_{21}(k,z) ~,
\label{eq:P21eq}
\end{equation}
where $\delta^D$ is the Dirac delta function, the brackets indicate average quantities, and $\widetilde{\delta}_{21}(\mathbf{k}, z)$ is the Fourier transform of ${\delta}_{21}(\mathbf{x}, z) = {\delta T}_{b}(\mathbf{x}, z)/ \overline{\delta T_b}(z) - 1$. In what follows, we shall work with the 21cm differential brightness temperature power spectrum, $\Delta^2_{21} (k,z) =(k^3/2 \pi^2) P_{21}(k,z)$, and show results for $\overline{\delta T_b}^2(z) \Delta^2_{21} (k,z)$ in \Sec{subsec:numer}.

\subsection{Local heating and ionization}
\label{sec:localPBH}

As mentioned in the previous section, within the ADAF framework only a small fraction of the energy gained via accretion is radiated into the medium. In this section we attempt to model the contribution to the 21cm line from the local environment of the PBH, estimating the radial profiles.

\subsubsection{Radial profiles}

The difficulty in estimating the true contribution to the 21cm signal from the local medium surrounding PBHs arises from the fact that solving for the radial density, $\rho$, velocity, $v$, temperature, $T$, and ionization profiles, $x_e$, of the accreting medium is non-trivial, and can only be solved analytically in very specific scenarios and certain regions of the parameter space (for an analytic solution at small radii within the ADAF framework, see, e.g., Ref.~\cite{Narayan:1994is}). As will be shown below, the dominant contribution to the 21cm signal comes from radii near the Bondi radius (the Bondi radius, to be defined, approximately demarcates where the solutions asymptote to the IGM value). This motivates studying the simple solution to the problem of static spherically symmetric accretion around a point, in which we neglect heating from the radiated emission (likely a valid approximation for the ADAF framework since the efficiency is small) and viscosity (an assumption which certainly brakes down near the BH, but which plays a less significant role as one approaches the IGM).
 
In this static spherically symmetric case, the fluid equations for determining the radial profiles are given by
\begin{align}
\label{eq:fluid1}
4\pi r^2 \, \rho \, |v| & = \dot{M}_{\rm PBH} = {\rm constant} ~,  \\
\label{eq:fluid2}
v \, \frac{dv}{dr} &= - \frac{1}{\rho} \, \frac{dP}{dr} -\frac{G M_{\rm PBH}}{r^2} ~, \\
\Lambda_{\rm ion} (1-x_e) &= \alpha_{\rm B} \, n_{\rm H} \, x_e^2 ~,
\label{eq:fluid3}
\end{align}
where $\alpha_{\rm B}$ is the case-B recombination coefficient, $P = \rho \, (1 + x_e) T/m_p$ is the pressure, and $m_p$ is the mass of the proton (we neglect the helium contribution in this computation for the sake of simplicity). The first of these equations arises from integrating the continuity equation, the second one is the Euler equation (accounting for momentum conservation), and the final one determines the balance between ionization and recombination. As in \Sec{sec:globalPBH}, we adopt the Bondi-Hoyle accretion rate given in \Eq{eq:BHrate}. The relevant scales are the Bondi velocity, defined as in \Eq{eq:veffparam}, 
and the Bondi radius
\begin{equation}
\label{eq:rBndi}
r_{\rm B} = \frac{G M_{\rm PBH}}{v_{\rm eff}^2} \, .
\end{equation}
Note that these definitions differ from those originally defined by Bondi, a consequence of the fact that we have modified the accretion rate via the substitution $c_s \rightarrow v_{\rm eff}$, as discussed above. For reference, a value of $v_{\rm eff} \sim c_s$ and $T_k \sim T_{\rm ad} \propto (1+z)^2$ lead to a Bondi radius of $r_{\rm B} \sim 3\times 10^{-5} \, (M_{\rm PBH}/M_\odot) (30/(1+z))^2$~kpc. For distances greater than the Bondi radius, the effects of accretion are negligible, asymptoting to their background values.

As mentioned in \Sec{sec:globalPBH}, the accretion of matter onto the central BH leads to a highly energetic radiation flux which is capable of ionizing the surrounding medium. In order to solve \Eq{eq:fluid3}, it is necessary to identify the flux of ionizing photons as a function of the radial distance to the BH. Here, we assume that most of the radiation is produced at very small radii (where the surrounding medium can reach temperatures in excess of $10^9$~K), and thus, one can write the flux in terms of the comoving distance coordinate $r$ from the source by using a point source approximation: 
\begin{equation}
F(E,r,z)=\frac{L_{\rm acc}(E, z)}{4\pi r^2}\, e^{-\tau_{\rm ion}(E,r,z)} ~, 
\end{equation}
where $L_{\rm acc}(E, z)$ is the accretion luminosity given by combining Eqs.~(\ref{eq:lacc}) and~(\ref{eq:L_spec}). The ionization optical depth $\tau_{\rm ion}$ is given by
\begin{equation}
\tau_{\rm ion}(E,r,z) = \int_0^r n_{\rm H}(z) \, \sigma(E) \, x_{\rm HI}(r,z) ~,
\end{equation}
where the ionization cross section $\sigma(E)$ is taken to be the fitting function obtained in Ref.~\cite{1989ApJ...344..551Z}. For most of the radii of interest, the optical depth is found to be negligible. The local ionization rate is given by
\begin{equation}
\Lambda_{\rm ion} = \int_{E_0}^{\infty} \frac{ dE}{E} \, \sigma(E) \, F(E,r,z) \left[ 1+ \frac{E - E_{\rm th}}{E_{\rm th}}f_{\rm ion}(E,x_e)\right] ~,
\end{equation}
where $E_{\rm th}$ is the ground level energy of the hydrogen. We parameterize the energy deposition function $f_{\rm ion}$\footnote{Note that this is conceptually the same as parameter defined in \Eq{eq:energydeposit}. However, in the case of local ionization, it is appropriate to apply the on-the-spot approximation, and thus the functional forms and dependencies of these parameters are different.} using different fitting formulas for $E\leq 0.5$~keV \cite{Dijkstra:2004ik} and $E>0.5$~keV~\cite{1985ApJ...298..268S}.

Since, at the scales of interest, we can safely neglect heating/cooling effects, we get the usual adiabatic relation between the temperature and the density $T \propto \rho^{2/3}$. The set of Eqs.~(\ref{eq:fluid1}--\ref{eq:fluid3}) can be easily solved numerically given the boundary conditions $\rho_\infty$ and $T_\infty$, which we take as the IGM values, that depend on both $M_{\rm PBH}$ and $f_{\rm PBH}$, and are calculated using the formalism presented in the previous section (\Sec{sec:globalPBH}). 

Is is important to bear in mind that after obtaining the temperature and density profiles, one must resolve the radiative transfer equation radially in order to obtain the differential brightness temperature. For these small enough length scales, the time evolution and expansion of the Universe can be neglected, leading to the following kinetic equation in terms of the brightness temperature at a frequency $\nu$, $T_{b,\nu}$:
\begin{equation}
\label{eq:radtrans}
\frac{dT_{b,\nu}(r)}{dr} = - \kappa_{\nu}(r)\left[T_{b,\nu}(r) - T_S(r)\right] ~, 
\end{equation}
with $\kappa_{\nu}$ being the absorption coefficient
\begin{equation}
\kappa_{\nu}(r) =\frac{h\nu_0}{4 \pi} \, \phi_{\nu} \left[ B_{01} \, n_0(r)  - B_{10} \, n_1(r)  \right] 
 \simeq \frac{3 c^2\, A_{10} }{32 \pi \, \nu_0^2} \, x_{\rm H}(r) \, n_{\rm H} \, \phi_{\nu} \, \frac{T_0}{T_S(r)} ~,
\end{equation}
where $B_{01}, B_{10}$, and $A_{10}$ are the Einstein coefficients between the triplet and singlet states, and $\phi_{\nu}$ is the line profile, given by the Doppler broadening
\begin{equation}
\phi_{\nu} = \frac{1}{\sqrt{\pi} \, \Delta \nu_{\rm D}}e^{- (\nu - \nu_0)^2/\Delta \nu_{\rm D}^2} ~,
\end{equation}
with $\Delta\nu_{\rm D} = \nu_0 \sqrt{2 \, T_k/m_p}$. 

The brightness temperature along a line of sight at a radial distance $r$ from the PBH, \Eq{eq:radtrans}, can be formally solved with the boundary condition at $r = 0$, leading to
\begin{equation}
T_{b,\nu}(r) = T_{b,\nu}(0) \, e^{-\tau_{\nu}(0,r)} + \int^{r}_{0}dr' \, \kappa_{\nu}(r') \, T_S(r') \, e^{-\tau_{\nu}(r',r)} ~,
\end{equation}
with the optical depth $\tau_{\nu}(r', r)$ defined as
\begin{equation}
\tau_{\nu}(r',r)=\int^r_{r'}dr'' \, \kappa_{\nu}(r'') ~.
\end{equation}
At distances near the PBH the medium would be fully ionized, leading to a vanishing $\kappa_{\nu}$. Therefore, it is appropriate to adopt the boundary condition $T_{b,\nu}(0) = T_{\rm CMB}$. Defining the differential brightness temperature as $\delta T_{b,\nu}(r) \equiv (T_{b,\nu}(r) - T_{\rm CMB})/(1+z)$, we get
\begin{equation}
\delta T_{b,\nu}(r)= \int^{r}_{0}dr' \, \kappa_{\nu}(r') \, \frac{T_S(r') - T_{\rm CMB}}{1+z} \, e^{-\tau_{\nu}(r',r)} ~.
\label{eq:T21prof}
\end{equation}
The spin temperature can be calculated using \Eq{eq:spinT}, with the collisional coupling coefficient, $y_k$, defined in \Eq{eq:yk} and the Lyman-$\alpha$ coefficient, $y_\alpha$, in \Eq{eq:yalpha}. The Lyman-$\alpha$ flux in $y_\alpha$ contains two contributions, $J_\alpha (r) = J_{\alpha}^{\rm loc}(r) + J_{\alpha}^{\rm back}$. The first of these comes from the local emission in the local medium, given by
\begin{equation}
J_{\alpha}^{\rm loc}(r) = \frac{f_{\alpha}(x_e) \, c}{4\pi \, H \, \nu_\alpha \, h\nu_\alpha} \, n_{\rm H}(z, r) \,  x_{\rm H}(r,z) \, \int_{E_0}^{\infty} \, dE \, \left(\frac{E - E_{\rm th}}{E}\right)\, \sigma(E) \, F(E,r,z) ~.
\end{equation}
We parameterize the energy deposition fraction $f_{\alpha}(x_e)$ using the fitting formula obtained in Ref.~\cite{1985ApJ...298..268S}, given by $f_{\alpha} = 0.48 \, (1 - x_e^{0.27})^{1.52}$. The second component to the Lyman-$\alpha$ flux comes from the background contribution, which is given the mean average over the IGM, produced by a combination of stellar sources and distant PBHs. This global flux $J_{\alpha}^{\rm back}$ for a given astrophysical model is computed by implementing the formalism of \Sec{sec:globalPBH} into {\tt 21cmFAST}, as discussed in the previous section. Thus, the Lyman-$\alpha$ coupling coefficient can then be determined summing both components.

\begin{figure}
	\centering
	\includegraphics[width=0.49\textwidth]{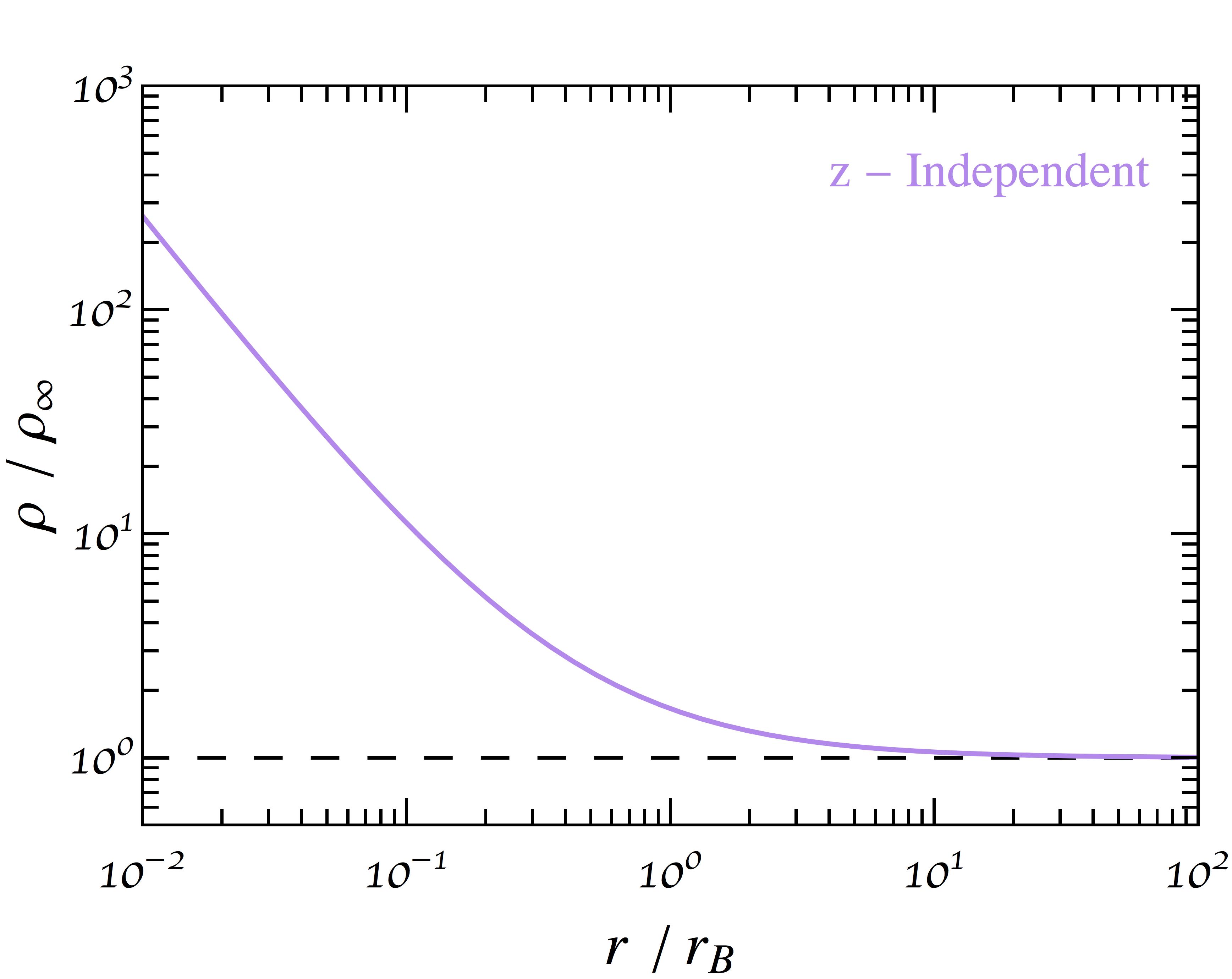}
	\includegraphics[width=0.49\textwidth]{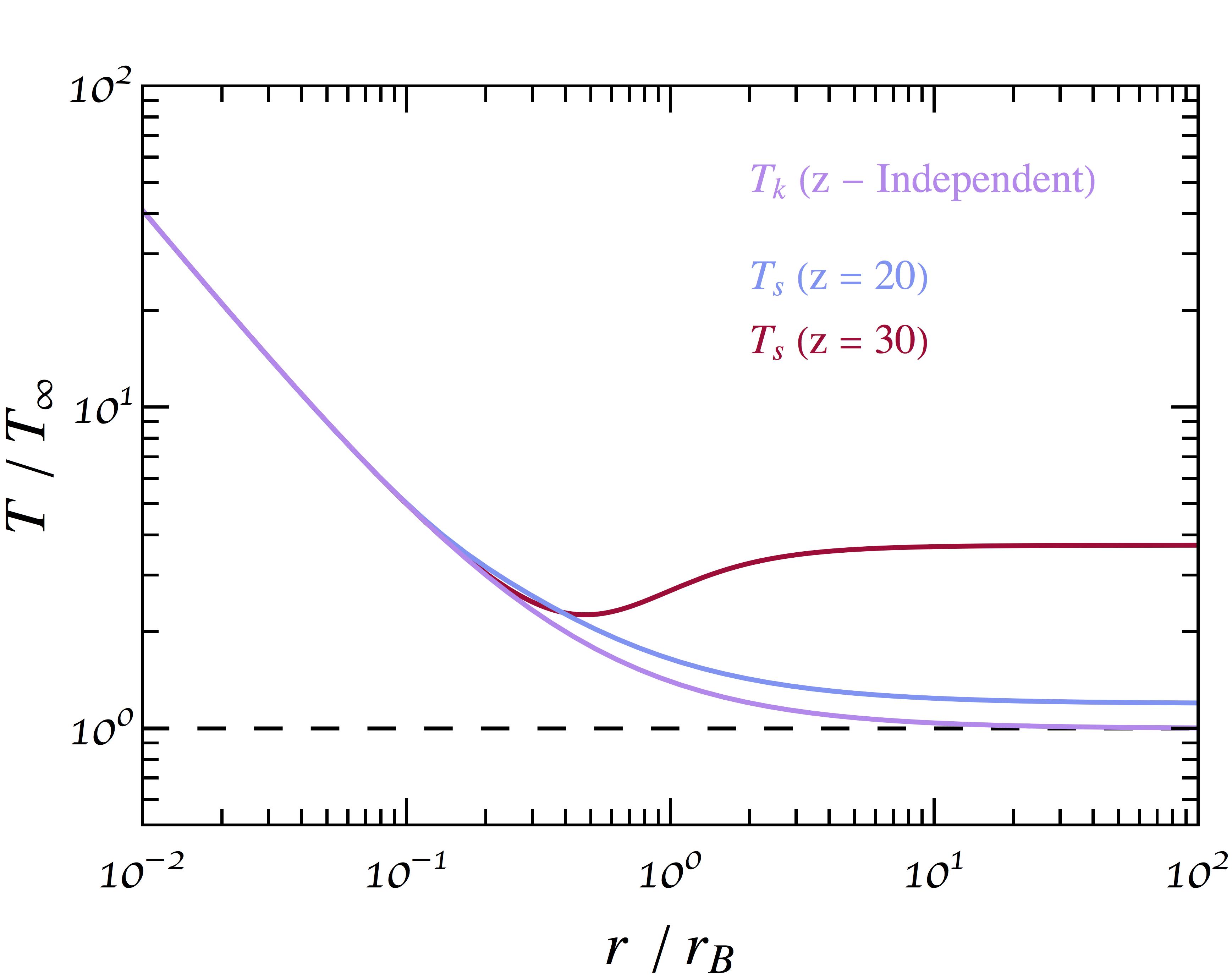}
	\includegraphics[width=0.49\textwidth]{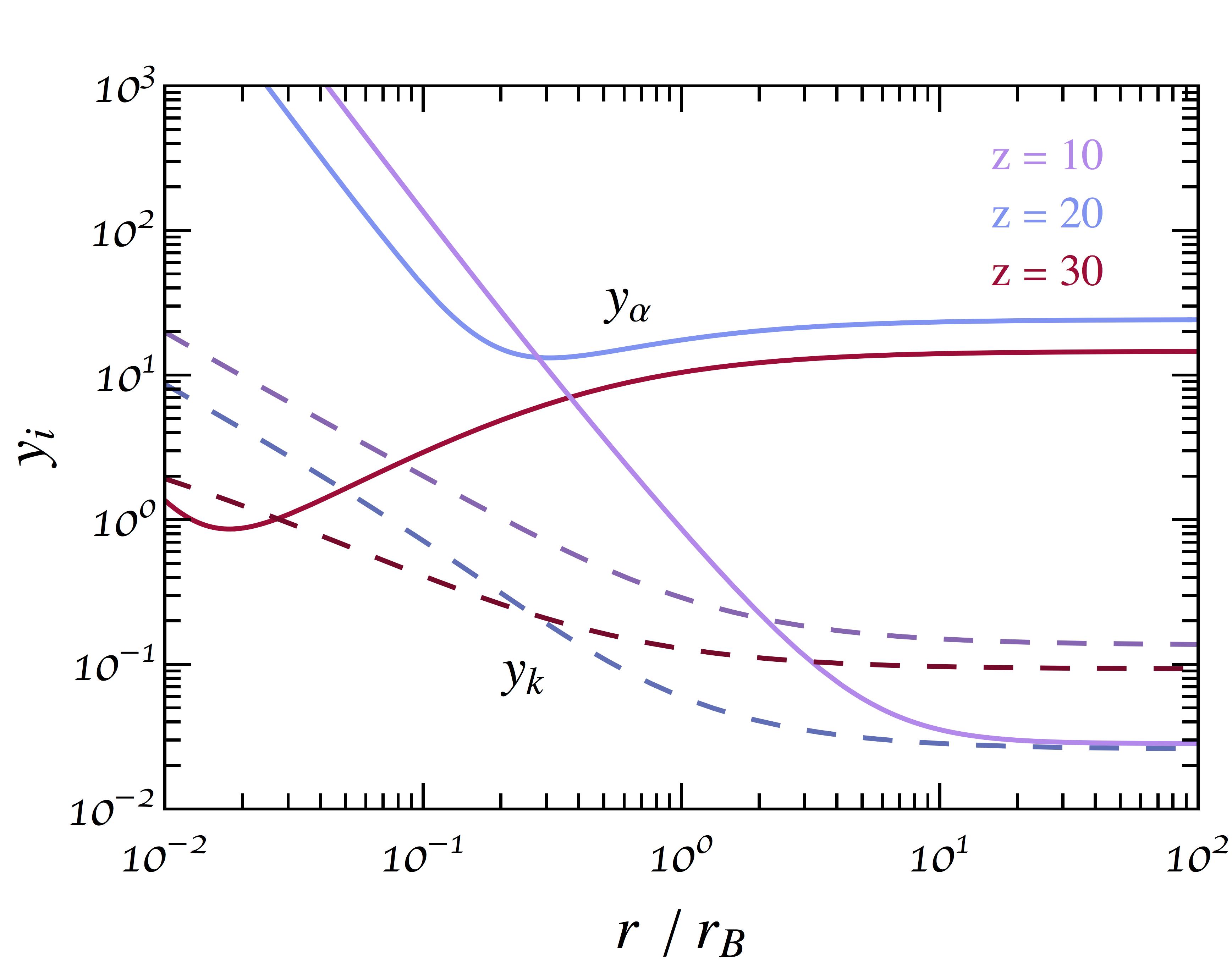}
	\caption{\label{fig:minihalos_profiles} Radial profiles for the local density (top-left panel), kinetic and spin temperatures (top-right panel), and coupling coefficients (bottom panel) $y_k$ (dashed lines) and $y_\alpha$ (solid lines), displayed at various redshifts, for a PBH of mass $M_{\rm PBH} = 1 \, M_\odot$. The dashed lines in the upper panels represent the IGM values.}
\end{figure}

\begin{figure}
	\centering
	\includegraphics[width=0.8\textwidth]{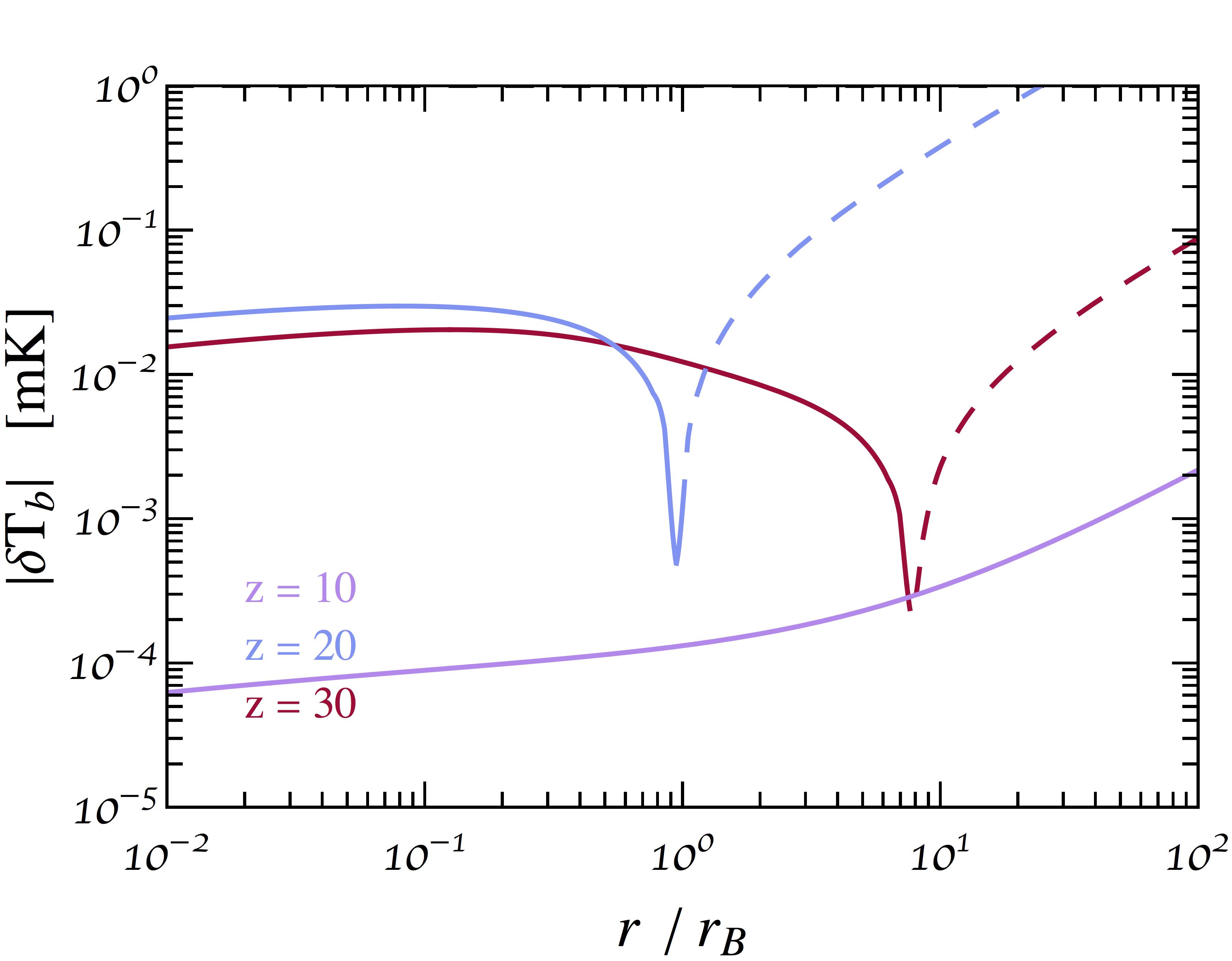}
	\caption{\label{fig:minihalo21cmP} Radial profile of the absolute value of the 21cm differential brightness temperature, $\delta \overline{T}_b$, for a PBH of mass $M_{\rm PBH} = 1 \, M_\odot$, at redshifts $z = 10$, $20$, and $30$. Dashed (solid) lines denote radii at which the signal would be seen in absorption (emission).}
\end{figure}

In \Fig{fig:minihalos_profiles}, we show radial profiles of the density (top-left panel), gas temperature (purple line in the top-right panel), spin temperature (blue and red lines in the top-right panel), and the collisional ($y_k$, solid lines in the bottom panel) and Lyman-$\alpha$ ($y_\alpha$, dashed lines in the bottom panel) couplings at various redshifts for $M_{\rm PBH} = 1 \, M_\odot$. As designed, both the density profiles and the kinetic temperature profiles asymptote to the IGM values, which have been taken from the customized version of {\tt 21cmFAST} (see \Sec{subsec:numer} for more details). The subsequent 21cm temperature profile, computed using the values in \Fig{fig:minihalos_profiles}, is depicted in \Fig{fig:minihalo21cmP}. Note that the contribution must be truncated at large radii to avoid double counting the IGM contribution --- we define this truncation to occur when the temperature profile is within $1\%$ of its asymptotic value, although the results are insensitive to a reasonable choice of this value. Note that the suppression of $y_\alpha$ (bottom panel in \Fig{fig:minihalos_profiles}) between $z = 20$ and $z = 10$ can be explained by its dependence with the inverse of the temperature of the gas, $y_\alpha \propto T_k^{-1}$, which has been already heated significantly by this epoch.

\newpage

\subsubsection{Averaged 21cm line signal}

The prescription provided above allows one to compute the 21cm profile for an individual PBH. Nevertheless, the observable quantity of interest is the contribution to the global 21cm line from all the PBHs in the Universe. Here, we re-derive the formula that allows one to compute the globally averaged brightness temperature $\overline{ T_b }$ from the radial 21cm profile~\cite{Iliev:2002gj}. The brightness temperature is related to the flux per unit frequency and solid angle $\frac{dF}{d\nu d\Omega}$ arising from a PBH of mass $M_{\rm PBH}$ by
\begin{equation}
\frac{dF}{d\nu d\Omega} = \frac{2\nu^2}{c^2}T_{b,\nu} ~.
\end{equation}
The averaged value of this quantity on a frequency interval $d\nu$ and a solid angle $d\Omega$ can be written as
\begin{equation}	
\overline{ \frac{dF}{d\nu d\Omega} } \, d\nu \, d\Omega = \langle F \rangle \, n_{\rm PBH} \, dV  ~,
\end{equation}
where $dV$ is the differential comoving volume, $n_{\rm PBH}$ is the number density of PBHs, and $\langle F \rangle$ is the area-averaged flux for a single PBH:
\begin{equation}
\label{eq:aa}
\langle F \rangle = \frac{1}{A} \int dA \, F ~,
\end{equation}
with $A = \pi r_{\rm max}^2$, and where $r_{\rm max}$ is the maximum radius at which the PBH influences the IGM (defined as above --- specifically, this is the radius at which $T_k$ is within $1\%$ of the asymptotic value). Using the fact that $|d\nu/dz|=\nu_0/(1+z)^2$, and writing the differential comoving volume as $dV = D^2dDd\Omega$, where $D$ is the comoving distance, we find
\begin{equation}
\overline{  \frac{dF}{d\nu d\Omega} } = \frac{c \, (1+z)^2}{\nu_0 \, H(z)} \, D^2 \, \langle F \rangle \, n_{\rm PBH}~,
\label{eq:df1}
\end{equation}
where we have used $dD/dz = c/H(z)$. The integrated flux over frequency and solid angle is then given by
\begin{equation}
F = \int d\nu \, d\Omega \, \frac{dF}{d\nu d\Omega} =  \Delta\Omega \, \Delta \nu_{\rm eff} \left. \frac{dF}{d\nu d\Omega} \right|_{\nu_0} ~, 
\end{equation}
where $\Delta\Omega = A/D_A^2$ is the solid angle subtended by the halo around the PBH, with $D_A=D/(1+z)$ the angular diameter distance, and $\Delta \nu_{\rm eff} = ((1+z) \, \phi_\nu(\nu_0))^{-1} = \sqrt{\pi} \, \Delta \nu_{\rm D}/(1+z)$~\cite{Iliev:2002gj}. One can then revert to the averaged brightness temperature using 
\begin{equation}
\overline{  \frac{dF}{d\nu d\Omega} } = \frac{2 \, \nu_0^2}{c^2} \, \overline{T}_b ~.
\label{eq:df2}
\end{equation}
From Eqs.~(\ref{eq:df1}) and~(\ref{eq:df2}), we find the globally averaged brightness temperature to be given by
\begin{equation}
\overline{T}_b = \frac{c \, (1+z)^4}{\nu_0 \, H(z)} \, n_{\rm PBH} \,\left\langle \Delta \nu_{\rm eff} \,  T_{b,\nu_0} \right\rangle A ~.
\label{eq:ilieveq}
\end{equation}

\begin{figure}
	\centering
	\includegraphics[width=0.49\textwidth]{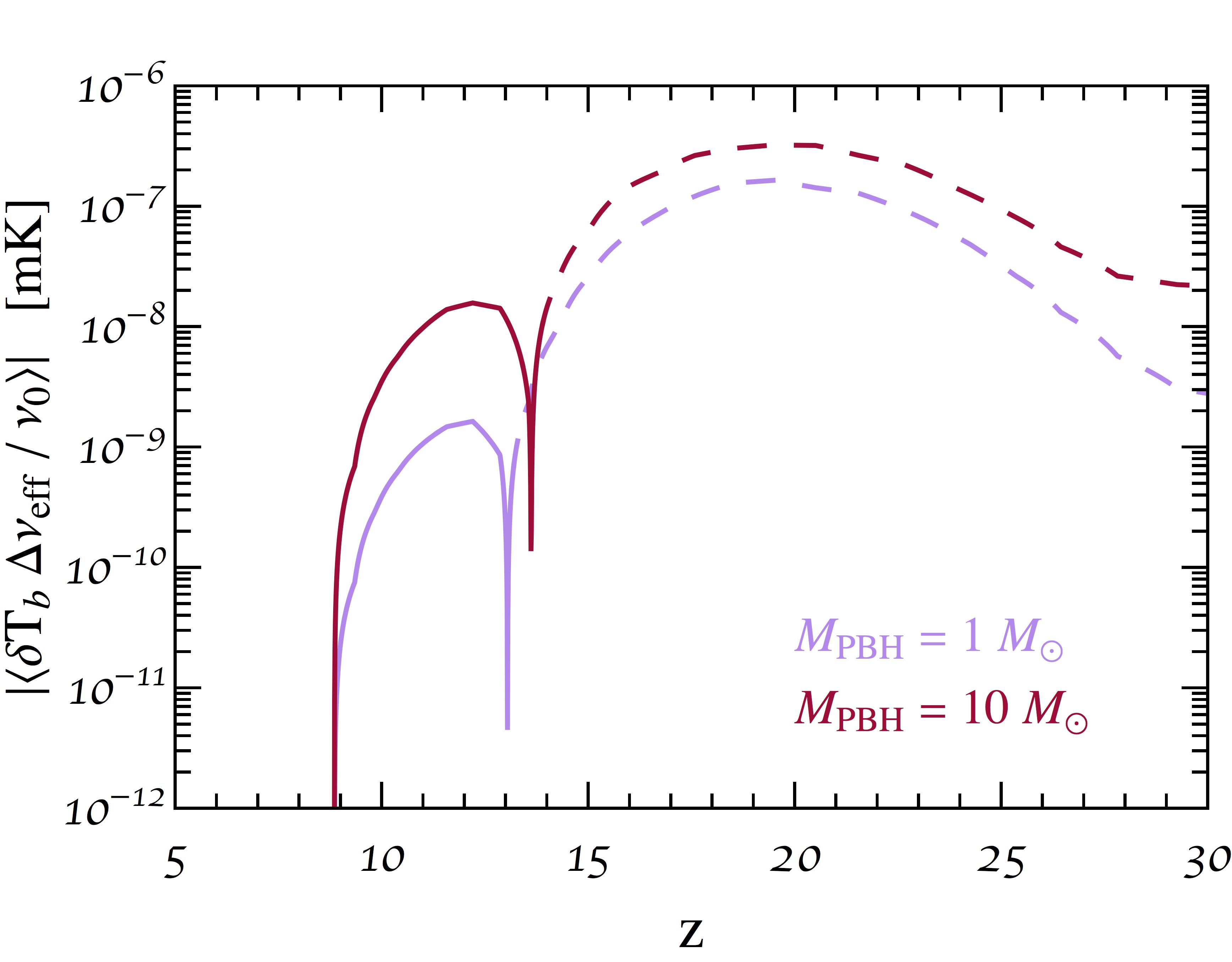}
	\includegraphics[width=0.49\textwidth]{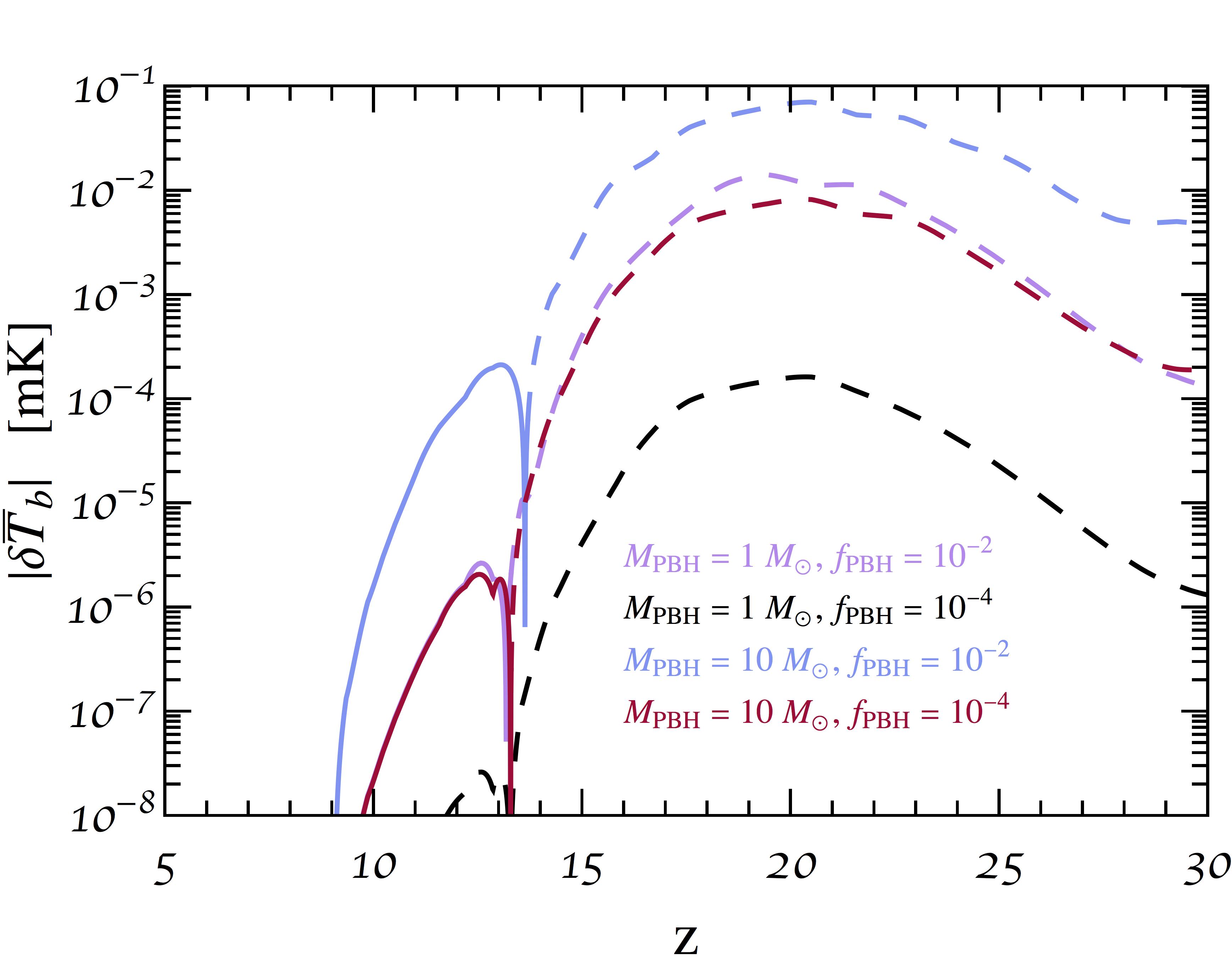}
	\caption{\label{fig:minihalos_averaged} Left Panel: Area-averaged differential brightness temperature times $\Delta\nu_{\rm eff}/\nu_0$ for one single PBH of mass $M_{\rm PBH} = 1 \, M_\odot$ and $M_{\rm PBH} = 10 \, M_\odot$, and assuming $f_{\rm PBH} = 10^{-2}$. Right Panel: Cumulative local contributions of a monochromatic population of PBHs to the globally averaged brightness temperature for mass $M_{\rm PBH} = 1 \, M_\odot$ and $M_{\rm PBH} = 10 \, M_\odot$, and dark matter fractions $f_{\rm PBH} = 10^{-2}$ and $f_{\rm PBH} = 10^{-4}$. In both panels, dashed (solid) lines represent when the 21cm signal would be seen in absorption (emission), i.e., $\delta T_b < 0$ ($\delta T_b > 0$).}
\end{figure}

For the area-averaged differential brightness temperature $\delta \overline{T}_b$, Eq.~(\ref{eq:ilieveq}) is applied but replacing $T_{b,\nu_0}$ by $\delta T_{b,\nu_0}$ computed from Eq.~(\ref{eq:T21prof}). We depict in \Fig{fig:minihalos_averaged} the area-averaged 21cm signal for one single PBH as a function of redshift (left panel), and the globally averaged signal (right panel), for various values of $M_{\rm PBH}$ and $f_{\rm PBH}$. The dashed and solid lines in the left panel indicate when the signal would be seen in absorption or emission. Interestingly, and perhaps counter-intuitively, the global contribution is often negative (i.e., it is seen in absorption). At first glance this appears strange, as the process of accretion heats and ionizes the medium --- thus, one might naively expect to see this contribution in emission. However, the asymptotic signal is observed in absorption during this epoch and, while the process of accretion does heat the local medium, the density rises faster than the temperature (recall that $\rho(r) \propto T(r)^{3/2}$). Thus, there exists a range of radii near the Bondi radius for which the kinetic temperature has not yet been heated above the background temperature, but for which the density has noticeably increased --- the net effect is an amplification of a slightly suppressed absorption dip. 

The large drop in the signal for $z\lesssim10$ in \Fig{fig:minihalos_averaged} appears because stellar sources have begun ionizing the medium. We emphasize that for the masses and dark matter fractions adopted here, the local contribution is never greater than $0.1$~mK. Given that the usual IGM signals for these epochs range between $1$ and $100$~mK (see section~\ref{subsec:numer}), the local contribution from PBHs is negligible for the models of interest.

\subsubsection{Comment on the local contribution}

Before continuing, a discussion on the treatment and impact of the local contribution is in order. We begin by addressing a number of concerns regarding the formalism adopted in the section above.

It should be emphasized that the spherically symmetric treatment of accretion adopted here neglects a number of potentially important features. First, our formalism only accounts for the  adiabatic evolution of the medium, leading to the relation $T \propto \rho^{2/3}$. Thus, we have neglected the local heating of the gas by the highly energetic radiation generated during accretion. This heating term is proportional to the luminosity, and therefore, to the radiative efficiency and the PBH mass. However, the effect of such a heating would be the enhancement of the local temperature, which would suppress the absorption signal that arises near the Bondi radius, subsequently reducing the global contribution to the 21cm signal. Thus, from this point of view, neglecting the heating term is a conservative assumption. We have also chosen to neglect viscosity in our treatment, despite the fact that viscosity is known to play an important role in ADAF accretion \cite{Xie:2012rs, Narayan:1994xi}. We believe that this choice is justified as the derived signal seems to be dominated by the behavior at larger radius, where the effects of viscosity are reduced. Nevertheless, a more rigorous numerical treatment would be required to verify this statement. 

Another concern about the adopted formalism is the assumption of the static solution of an isolated PBH in a uniform background. In reality, this is almost certainly not the case. The matter accreted onto the PBH must come from the IGM. We have implicitly assumed, however, that the asymptotic condition of the IGM remains unaltered by the presence of the PBH. Conceptually, one would expect that an underdensity at $r \gtrsim r_{\rm B}$ is necessary in order to counteract the overdensity observed for $r \lesssim r_{\rm B}$. Neglecting momentarily the effects of heating and ionization, the contribution of the underdensity should identically cancel that of the overdensity. The fact that the signal is predominately seen in absorption indicates that it is dominated by the state of the IGM near the Bondi radius, where not properly accounting for these changes in density could be important. Thus, it is entirely possible that the observed absorption feature is actually artificial, and arises from the fact that we have not correctly accounted for the influence of the PBH on the IGM. Nevertheless, this implies that our formalism is likely to significantly overestimate the true contribution of local accretion, again suggesting the estimate is conservative. 

Before continuing, it is important to address why the results found here differ by orders of magnitude from previous PBH calculations~\cite{Tashiro:2012qe, Bernal:2017nec}. We think there are multiple reasons for this discrepancy,\footnote{In addition to the differences in formalism, discussed below, the de-excitation rates (taken from Ref.~\cite{Kuhlen:2005cm}) appear to be incorrectly transcribed in Ref.~\cite{Bernal:2017nec}. Most notably, the \textit{logarithms} in the e-H de-excitation rate should be base 10, rather than natural logs. We have verified, using the formalism of Ref.~\cite{Bernal:2017nec}, that this mistake does indeed enter the calculations. Implementing the correct de-excitation rate into the same formalism completely removes the absorption dip responsible for the majority of the signal (see Fig.~6 of Ref.~\cite{Bernal:2017nec}), and thus significantly alters the contribution of minihalos to the global 21cm signal. The check we have performed agrees with the result of Ref.~\cite{Tashiro:2012qe}, which appears to have a formalism nearly identical to that of Ref.~\cite{Bernal:2017nec}, but without the artificial absorption dip.} which we outline below.

First, note that the assumed PBH luminosity in Ref.~\cite{Bernal:2017nec} is near the Eddington limit (specifically they adopt a luminosity of $L =  0.1 \, L_{\rm Edd}$). Such a large luminosity is not consistent with ADAF accretion, but rather solutions in which one forms a thin disk that radiates extremely efficiently ($\epsilon \sim 1$). In the event that only the radiative efficiency is modified, local heating would be reduced, and global heating enhanced.  However, there also exists a significant difference in the normalization of the luminosity, as the dimensionless accretion rate adopted in our framework is roughly $L/L_{\rm Edd} \sim 10^{-10}$. The difference in adopted luminosity changes the relative importance of heating, implying that within our formalism one cannot neglect the derivative terms in the accretion equations, whereas for large luminosities these terms may be subdominant to local heating.  Next, it is slightly concerning that Ref.~\cite{Bernal:2017nec} seeks a static solution to the local profile on scales comparable to and higher than the Hubble length (note that $r_{\rm H} = c/H(z)\simeq 4.6 \times 10^4  \left[ 30/(1+z) \right]^{3/2}$~kpc). Hence, the expansion of the Universe can no longer be neglected, and thus the static framework is applied out of the range of validity. This does not appear to be a problem with the ADAF framework, as the smaller luminosities imply smaller scales of interest.

\begin{figure}
\includegraphics[width=0.8\textwidth]{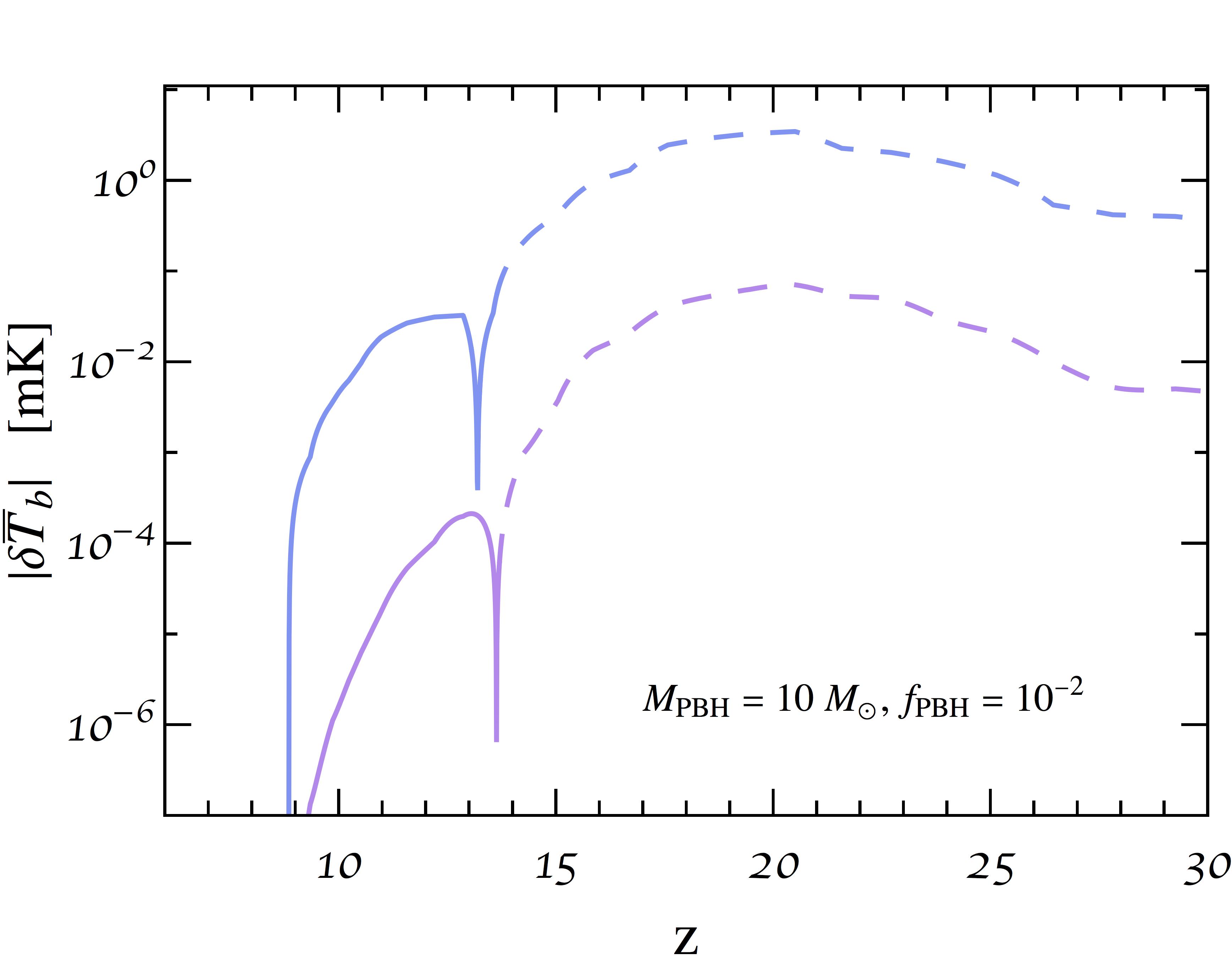}
\caption{\label{fig:incorrectTB} Contribution to differential brightness temperature, $\delta \overline{T}_b$, from local heating induced by accretion onto a population of PBHs with $M_{\rm PBH} = 10 \, M_\odot$,  assuming $f_{\rm PBH} = 10^{-2}$, computed by resolving the radiative transfer equation locally, \Eq{eq:T21prof}, (purple line) and by incorrectly applying \Eq{eq:Tbdev} (blue line). }
\end{figure}

Finally, it is important to note that one should not directly apply \Eq{eq:Tbdev} to obtain the brightness temperature profile around the PBH. The \Eq{eq:Tbdev} is only valid when considering large scales; namely it is valid when one can neglect the radial dependence of both the spin temperature and the optical depth, and when the optical depth is exclusively given by the optical depth of the IGM. Here, neither of these statements are true, and thus one must solve the radiative transfer equation as a function of the radial coordinate using  \Eq{eq:T21prof} (as was done in previous works, see, e.g., Refs.~\cite{Iliev:2002gj, Gong:2017sie}). This profoundly impacts the strength of the signal. In \Fig{fig:incorrectTB} we illustrate, using the formalism of the previous section, that incorrectly applying \Eq{eq:Tbdev} leads to an enhancement of the signal by roughly $\sim2$ orders of magnitude.

Therefore, we conclude that it is unlikely that the local contribution from PBHs could significantly add to the 21cm signal. Nevertheless, it is likely the case that complex high-resolution simulations will need to be performed for this contribution to be truly understood.

\subsection{Shot noise}
\label{sec:shot_N}

Since PBHs are point sources, they contribute to the power spectrum as Poisson white noise (i.e., they contribute in a scale-independent manner)~\cite{Afshordi:2003zb}. The associated perturbations are isocurvature modes and thus, their contribution only affects scales smaller than those corresponding to the matter-radiation equality era, $k > k_{\rm{\rm eq}} \equiv a_{\rm eq} H_{\rm eq}$, being negligible at larger scales. Approximating the isocurvature transfer function as $T_{\rm iso} = 3/2 \, (1+z_{\rm eq})$ for $k>k_{\rm{\rm eq}}$ and $0$ otherwise, where $z_{\rm{eq}}$ is the redshift of matter-radiation equality~\cite{Peacock:1999ye}, this term can be written as~\cite{Afshordi:2003zb}
\begin{equation}
\label{eq:Pnoise}
\Delta P = T^2_{\rm iso} \, \frac{f_{\textrm{PBH}}^2}{n_{\textrm{PBH}}} \, D^2(z)
= \frac{9 \, (1+z_{\rm{eq}})^2 f_{\textrm{PBH}} \, M_{\rm PBH}}{4 \, \Omega_{\rm DM} \, \rho_c} D^2(z) \simeq 2.5 \times 10^{-2} f_{\textrm{PBH}} \left(\frac{M_{\rm PBH}}{30 \, M_\odot}\right) D^2(z)\ \textrm{Mpc}^3~,\quad k>k_{\rm{eq}} ~,
\end{equation}
where $\Omega_{\rm DM}$ is the fraction of dark matter energy density over the critical density $\rho_c$ and $D(z)$ is the linear growth factor of density perturbations. The inclusion of this contribution modifies the halo mass function on small scales. For the masses studied here (namely $M_\odot \lesssim M_{\rm PBH} \lesssim 10^3 M_\odot$), however, the effects are only important for halos which are not large enough to cool and collapse to form stars (such halos are known as `minihalos'). Since star-forming halos are expected to produce the dominant 21cm signal in the epochs of interest, one is naively driven to conclude that the Poisson noise in the power spectrum is not relevant and can be neglected (even though, we have included it in the calculation of the global signal). It was recently argued~\cite{Gong:2017sie, Gong:2018sos}, however, that the enhanced number density of minihalos arising from the shot noise induced by $\sim \mathcal{O}(M_\odot)$ PBHs could sufficiently elevate the contribution of minihalos to a discernible level. We briefly outline the formalism for estimating the minihalo contribution below, and show that the computation of Ref.~\cite{Gong:2017sie} failed to self-consistently account for the heating of the IGM induced by PBH accretion. Correctly accounting for this effect tends to suppress the minihalo contribution to the point where it can be neglected.

\subsubsection{Brightness temperature inside minihalos}

A dark matter halo at redshift $z$ can be characterized by one parameter, the overdensity $\Delta_c(z)$ relative to the critical density $\rho_c(z)$, so that its mass $M_{\rm h}$ and virial radius $R_{\rm vir}$ are related by
\begin{equation}
M_{\rm h} = \frac{4\pi}{3} \, \Delta_c(z) \, \rho_c(z) \, R_{\rm vir}^3 ~.
\label{eq:Rvir}
\end{equation}
Here, we use the virial overdensity based on spherical collapse~\cite{Bryan:1997dn}
\begin{equation}
\Delta_c(z) = 18\pi^2 + 82 \, (\Omega_{\textrm{m},z} - 1) - 39 \, (\Omega_{\textrm{m},z} - 1)^2 ~,  
\end{equation}
where 
\begin{equation}
\Omega_{\textrm{m},z} = \frac{\Omega_{\rm m} \, (1+z)^3}{\Omega_{\rm m} \, (1+z)^3 + \Omega_\Lambda} ~.
\end{equation}
In the calculation below, we assume the neutral hydrogen fraction, $x_{\rm H}$, to be given by the IGM value, and the kinetic temperature in the halo by $T_k = T_{\rm vir}$, where the virial temperature of a truncated isothermal halo is~\cite{Shapiro:1998zp} 
\begin{equation}
T_{\rm vir}(M_{\rm h}, z) \simeq 4.8 \times10^4 \, \textrm{K} \, \left(\frac{\mu}{1.22} \, \right) \left(\frac{M_{\rm h}}{10^8 M_\odot \, h^{-1}} \, \right)^{2/3} \left(\frac{\Omega_{\rm m}}{\Omega_{\textrm{m},z}} \, \frac{\Delta_c(z)}{18\pi^2} \right)^{1/3} \left(\frac{1+z}{10}\right)  ~,
\label{eq:Tvir}
\end{equation}
being $\mu  = 1.22$ the mean molecular weight of neutral IGM. The average number density of neutral hydrogen of the halo is
\begin{equation}\label{eq:meannh}
\bar{n}_{\rm H}(M_{\rm h},z) = \frac{\mathfrak{f}_{\rm H}}{m_p} \, \left(\frac{\Omega_{\rm b}}{\Omega_{\rm m}}\right) \, \left(\frac{3 \, M_{\rm h}}{4 \pi \, R_{\rm vir}^3}\right) = \mathfrak{f}_{\rm H} \,  \left(\frac{\Omega_{\rm b}}{\Omega_{\rm m}}\right) \, \left(\frac{\rho_c(z) \, \Delta_c(z)}{m_p}\right) ~,
\end{equation}
where $\rho_b$ the baryon energy density. For the sake of consistency with the constant temperature assumption, we consider that neutral hydrogen in halos follows a truncated (singular) isothermal distribution, normalized to provide the mean value given in \Eq{eq:meannh},
\begin{equation}
n_{\rm H}(M_{\rm h},z,r) = \frac{1}{3} \bar{n}_{\rm H}(M_{\rm h},z) \left( \frac{R_{\rm vir}}{r} \right)^2 ~.
\end{equation}
We have also verified that reasonable modifications to this assumption (e.g., taking an NFW profile~\cite{Navarro:1996gj} or a truncated non-singular isothermal profile~\cite{Shapiro:1998zp}) give rise to only small changes on the cumulative minihalo signal.

Since the flux of Lyman-$\alpha$ photons predominantly arises from star formation, the spin temperature \emph{inside} minihalos, whose virial temperatures are below threshold for atomic hydrogen cooling, is dictated exclusively by the collisional coupling, $y_k$, given by \Eq{eq:yk}. The brightness temperature for photons with frequency $\nu$ through a minihalo of given mass and redshift is obtained by solving the radiative transfer equation. Here, however, one has to be careful so as to not double count factors contained in the IGM contribution. To clarify, consider decomposing the IGM and minihalo contribution to the spin temperature such that  
\begin{eqnarray}
\label{eq:mh_tau_list}
\tau & = & \tau_{\rm IGM} + \tau_{\rm mh} ~, \\
T_S & = & T_{S, \rm IGM} + T_{S, \rm mh} ~, 
\end{eqnarray}
where $\tau_{\rm IGM}$ and $\tau_{\rm mh}$ are the optical depth of the IGM and the minihalo, respectively, and we define
\begin{eqnarray}
\tau_{\rm mh}, \, T_{S, \rm mh} & \neq 0 & \hspace{.4cm} {\rm for} \hspace{.3cm} |r| \leq R_{\rm vir}  ~, \\
\label{eq:mh_ts_list}
\tau_{\rm IGM}, \, T_{S, \rm IGM} & \neq 0 & \hspace{.4cm} {\rm for} \hspace{.3cm} |r| > R_{\rm vir}  ~,
\end{eqnarray}
where $r$ is the radial distance from the center of a minihalo with maximum radius $R_{\rm vir}$. The general solution to the radiative transfer equation (neglecting scattering) is given by
\begin{equation}
T_b = T_{\rm CMB} \, e^{-\tau} + \int \, d\tau \, T_S(\tau) \, e^{-\tau}  ~.
\end{equation}
Using Eqs.~(\ref{eq:mh_tau_list}--\ref{eq:mh_ts_list}), one finds
\begin{equation}
T_b = T_{\rm CMB} \, e^{-(\tau_{\rm IGM}+\tau_{\rm mh} )} + \int \, ds \, \frac{d\tau_{\rm IGM}}{ds} \, T_{S, \rm IGM} \, e^{-(\tau_{\rm IGM}+\tau_{\rm mh} )} + \int \, ds \, \frac{d\tau_{\rm mh}}{ds} \, T_{S, \rm mh} \, e^{-(\tau_{\rm IGM}+\tau_{\rm mh} )}  \, ,
\end{equation}
where $s$ is the line-of-sight. In order to avoid double counting, one must subtract off the contribution of the IGM, given by
\begin{equation}
T_{b, \rm IGM} =  T_{\rm CMB} \, e^{-\tau_{\rm IGM}} + \int \, ds \, \frac{d\tau_{\rm IGM}}{ds} \, T_{S, \rm IGM} \, e^{-\tau_{\rm IGM}} \, ,
\end{equation}
implying the solution for the minihalo brightness temperature is given by
\begin{equation}
\label{eq:tb_mh}
T_{b, \rm mh}(\nu) = T_{\rm CMB} \, e^{-\tau_{\rm IGM}} \, (e^{-\tau_{\rm mh}(\nu,\infty)} - 1) + \int_{-\infty}^\infty \, ds \, T_S(\ell) \, e^{- \left(\tau_{\rm IGM} + \tau_{\rm mh}(\nu,s) \right)} \, \frac{d\tau_{\rm mh}(\nu,s)}{ds} ~, 
\end{equation}
where we have added some explicit dependences. The optical depth of the minihalo, $\tau_{\rm mh}(\nu,s)$, is given by
\begin{equation}
\label{eq:taumh}
\tau_{\rm mh}(\nu,s) = \frac{3 \, A_{10} \, T_0}{32\pi \, \nu_0^2} \, \int_{-\infty}^s \, ds' \, \frac{n_{\rm H}(\ell) \, \phi(\nu,\ell)}{T_S(\ell)} ~, 
\end{equation}
and $s$ and $\ell$ are radial comoving distances, $s$ being the component along the line-of-sight and $\ell$ connecting $s$ to the center of the minihalo, $\ell^2 = s^2 + b^2 R_{\rm vir}^2$~\cite{Chongchitnan:2012we}, with $b$ the impact parameter. One might be inclined to obtain the differential brightness temperature by subtracting $T_{\rm CMB}$ from \Eq{eq:tb_mh}, however note that it is given by
\begin{equation}
\delta T_b = \frac{T_b - T_{\rm CMB}}{1 + z} = \frac{T_{b, \rm IGM} + T_{b, \rm mh} - T_{\rm CMB}}{1 + z} = \frac{\delta T_{b, \rm IGM} + T_{b, \rm mh}}{1 + z} ~,
\end{equation}
and thus the proper contribution to the total differential brightness temperature is simply given by \Eq{eq:tb_mh}.  

Obtaining the global contribution from all minihalos can be done in a similar manner to what was shown in \Sec{sec:localPBH}. Specifically, one can use \Eq{eq:ilieveq}, but with the generalization that $n_{\rm PBH} \rightarrow \int \, dM_{\rm h} \, dn/dM_{\rm h}$, i.e., 
\begin{equation}
\delta\overline{T}_b = \frac{c \, (1+z)^4}{\nu_0 \, H(z)} \int_{M_{\rm min}}^{M_{\rm max}} \, dM_{\rm h} \frac{dn}{dM_{\rm h}} \,\left\langle \Delta \nu_{\rm eff} \, \delta T_{b,\nu_0} \right\rangle A ~,
\label{eq:ilieveqmh}
\end{equation}
where, in analogy to the area-averaged signal in \Sec{sec:localPBH}, the brackets represent the average over the minihalo cross section (c.f., \Eq{eq:aa}), $A = \pi R_{\rm vir}^2$. The integration of \Eq{eq:ilieveqmh} is bounded from above by the mass threshold for star formation, which we take as the one corresponding to a virial temperature, \Eq{eq:Tvir}, of $T_{\rm min} = 10^4$~K. The lower limit of the integral is chosen as the maximum between the PBH mass and the Jeans mass $M_{\rm J}$ (although for all cases of interest here, $M_{\rm PBH} < M_{\rm J}$), which is given by the mass enclosed in a region with diameter of the Jeans length, $\lambda_{\rm J}=\sqrt{(5 \pi \, T_k)/(3 \, G \, \bar{\rho} \, \mu \, m_p)}$, 
\begin{equation}
M_{\rm J}(z) = \frac{4 \pi}{3}  \bar{\rho} \left( \frac{\lambda_{\rm J}}{2} \right)^3 \simeq 2.2 \times 10^9 M_\odot  \left(\frac{T_k}{10^4~{\rm K}}\right)^{3/2} \, \left(\frac{1.22}{\mu} \right)^{3/2} \, \left(\frac{10}{1+z}\right)^{3/2} ~,
\end{equation}
where $\bar{\rho}$ is the mean matter density prior to collapse. Here, we calculate $T_k$ using the formalism presented in~\Sec{sec:globalPBH}.

\begin{figure}
	\includegraphics[width=0.8\textwidth]{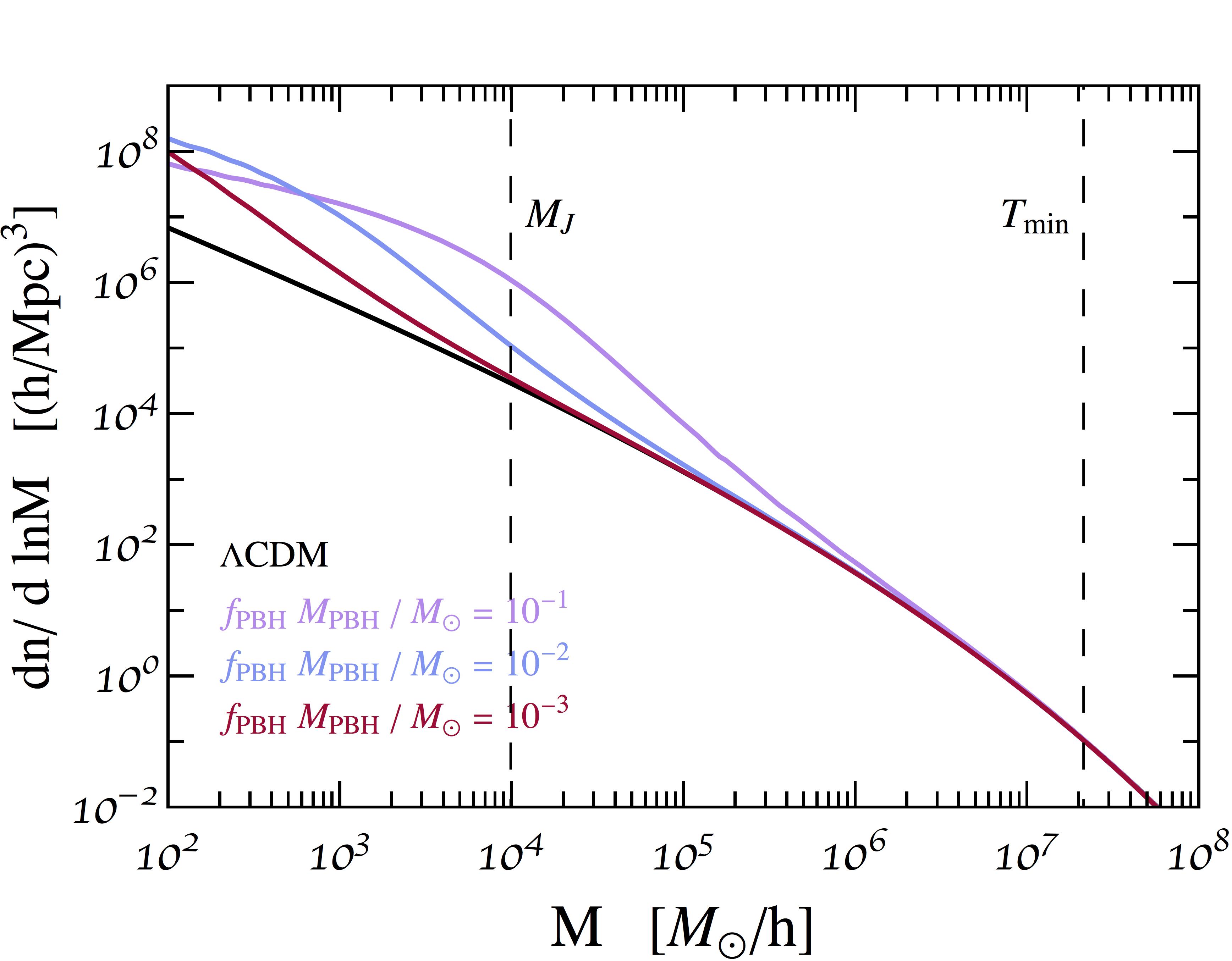}
	\caption{Halo mass function for several PBH cases at $z = 20$. Dotted and dashed vertical lines represent the Jeans mass, assuming a $\Lambda$CDM cosmology (i.e., not additional heating from PBH accretion), and the minimum halo mass for star formation, respectively, which limit the range of integration in \Eq{eq:ilieveqmh}.}
\label{fig:hmf_pbh}
\end{figure}

The number density of halos per unit mass is given by the halo mass function,
\begin{equation}
\label{eq:dndm}
\frac{dn}{dM_{\rm h}} = \frac{\rho_{\rm m}}{M_{\rm h}} \, \frac{d \ln \sigma^{-1}}{d M_{\rm h}} \, f(\sigma) ~,
\end{equation}
where $f(\sigma)$ represents the fraction of mass that has collapsed to form halos per unit interval in $\ln \sigma^{-1}$, with $\sigma$ the root-mean-square of the matter density fluctuation smoothed with a real-space top hat filter over the virial radius, 
\begin{equation}
\sigma^2(M_{\rm h}, z) = \frac{D^2(z)}{D^2(0)} \, \int_{0}^{\infty} \, \frac{dk}{k} \, \frac{k^3 \, P(k)}{2 \pi^2} \, |W(k; M_{\rm h})|^2 ~,
\end{equation}
where $D(z)$ is the growth factor of linear perturbations, $P(k)$ is the linear matter power spectrum (here, modified by including the shot noise term from the PBH population, \Eq{eq:Pnoise}), and $W(k; M_{\rm h})$ is the Fourier transform of the real-space top hat filter. Here we use the Sheth-Tormen halo mass function~\cite{Sheth:1999mn, Sheth:1999su},
\begin{equation}
f(\sigma) = A \, \sqrt{\frac{2 \, a}{\pi}} \, \left[ 1 + \left(\frac{\sigma^2}{a \, \delta_c^2} \right)^p\right]\frac{\delta_c}{\sigma} \, e^{-\frac{a \delta_c^2}{2 \sigma^2}} ~,
\end{equation}
with fit coefficients $A = 0.3222$, $a = 0.707$, and $p = 0.3$, and the critical collapse density $\delta_c = 1.686$. The halo mass function at $z = 20$ for several PBH cases is depicted in \Fig{fig:hmf_pbh}. Note that, since the additional shot noise contribution from PBH to the matter power spectrum depends on the combination $f_{\rm PBH}M_{\rm PBH}$ (see \Eq{eq:Pnoise}), so does the modification of the halo mass function with respect to the standard $\Lambda$CDM case.

\begin{figure}
	\includegraphics[width=0.8\textwidth]{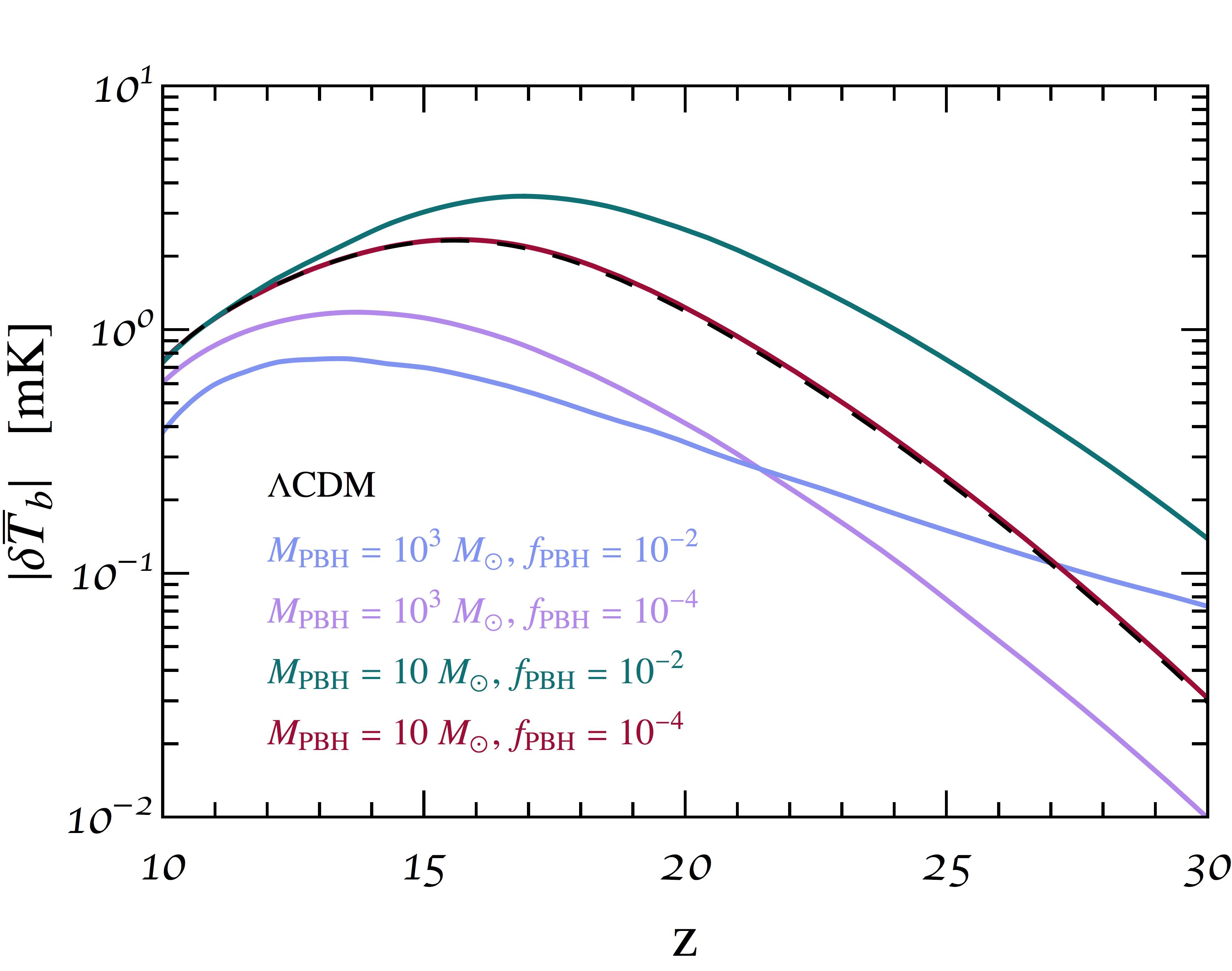}
	\caption{\label{fig:minihalo_PS} Collective contribution of minihalos to the globally averaged 21cm differential brightness temperature, $\delta \overline{T}_b$, for various PBH masses and dark matter fractions, and compared with the minihalo contribution in $\Lambda$CDM (black dashed line), as a function of redshift.}
\end{figure}

Using Eqs.~(\ref{eq:tb_mh}), (\ref{eq:ilieveqmh}), and (\ref{eq:dndm}), one can calculate the globally averaged 21cm signal from the cumulative population of minihalos. In \Fig{fig:minihalo_PS}, we depict the contribution of minihalos assuming a monochromatic population of PBHs with mass of either $10~M_\odot$ or $10^3~M_\odot$ and a dark matter fraction of $f_{\rm PBH} = 10^{-2}$ or $f_{\rm PBH} = 10^{-4}$. Notice that the redshift dependence of the minihalo contribution can be rather non-trivial, as the $T_k$ dependence of the Jeans mass is modified by the accretion model, and the mass derivative of the halo mass function can be either enhanced or suppressed relative to $\Lambda$CDM at this mass scale. We can see that the enhancement on the formation of structures due to the Poisson noise would produce a larger signal in the case $M_{\rm PBH} = 10~M_\odot$ and $f_{\rm PBH} = 10^{-2}$. However, contrary to previous claims~\cite{Gong:2017sie, Gong:2018sos}, the other cases shown in \Fig{fig:minihalo_PS} are not enhanced but suppressed. For small PBH masses and low values of $f_{\rm PBH}$, the minihalo contribution saturates to the standard $\Lambda$CDM prediction; that is to say, the number density of minihalos is negligibly modified, and the result is equivalent to that of $\Lambda$CDM. Meanwhile, the signal is suppressed for models with larger values of $M_{\rm PBH} \times f_{\rm PBH}$, where the enhancement of the halo mass function is significant. The reason for the discrepancy between the result presented here and that found in Refs.~\cite{Gong:2017sie, Gong:2018sos} lies on the treatment of the Jeans mass. In those previous works, only adiabatic cooling was considered for the evolution of the IGM temperature, while here the heating by standard X-ray sources and by accretion onto the population of PBHs (see \Sec{sec:globalPBH}) has also been included. This implies a larger Jeans mass and thus, a reduction of the relevant mass range of integration in \Eq{eq:ilieveqmh}. For instance, for the models depicted in \Fig{fig:Tk}, X-ray radiation would heat up the temperature around an order of magnitude with respect to the adiabatic cooling case at $z \sim 10$, increasing therefore the Jeans mass by a factor of $10^{3/2} \sim 30$.

\section{Methodology }
\label{sec:methods}

\subsection{Numerical simulations}
\label{subsec:numer}

As already mentioned above, in order to compute the redshift evolution of both the 21cm global signal and power spectrum we make use of the publicly available package {\tt 21cmFASTv1.2}~\cite{Mesinger:2010ne}. This code produces realizations of the evolved density, ionization, peculiar velocity and spin temperature fields from semi-analytic calculations. It depends on a number of parameters describing the different processes taking place at the reionization period. We use here a minimal set of four parameters: the ultraviolet (UV) ionization efficiency, $\xi_{\rm UV}$, the number of X-ray photons per solar mass, $\xi_{\rm X}$, the minimum virial temperature of halos hosting galaxies, $T_{\rm min}$, and the number of photons per stellar baryon between Lyman-$\alpha$ and the Lyman limit, $N_\alpha$.~\footnote{The default value of $N_{\alpha}$ in {\tt 21cmFAST} is obtained by assuming Pop II stars~\cite{Barkana:2004vb} and normalizing their emissivity to $\sim 4400$ ionizing photons per stellar baryon.} The phenomenological parameter $\xi_{\rm UV}$ is assumed to be proportional to: \textit{(i)} the fraction of ionizing photons escaping their host galaxies, \textit{(ii)} the number of ionizing photons per stellar baryons inside stars, and \textit{(iii)} the fraction of baryons that form stars. The efficiency for ionization, heating, and Lyman-$\alpha$ production by X-ray sources depends on the total X-ray emission rate, which is proportional to the star formation rate and to the number of X-ray photons per solar mass in stars, $\xi_{\rm X}$.\footnote{A value of $\xi_{\rm X} = 10^{56} \, M_\odot^{-1}$ implies $N_{\rm X} \simeq 0.1$ X-ray photons per stellar baryon.} The temperature $T_{\rm min}$ is the minimum of the virial temperature of a halo, below which gas cannot cool efficiently, and thus star formation is suppressed.\footnote{Note that the definition of the virial temperature here is taken to be as defined in Ref.~\cite{Barkana:2000fd}, which differs from \Eq{eq:Tvir} by a factor of $2.44$. This difference arises from the fact that in \Eq{eq:Tvir} we consider $\mu = 1.22$ and that the virial temperature depends upon the adopted halo profile, although these changes do not significantly impact any of the results of this section. The default value in the {\tt 21cmFAST} code is $T_{\rm min} = 10^4$~K, which has been identified in the literature with the atomic cooling threshold~\cite{Evrard:1990, Blanchard:1992, Tegmark:1996yt, Ciardi:1998zx, Haiman:1999mn}, and using Ref.~\cite{Barkana:2000fd}, it corresponds to a minimum halo mass of $M_{\rm min} \simeq 8 \times 10^{7} \, M_\odot$ at a redshift $z = 10$.} In the following, we present various examples of the 21cm signal using a fiducial $\Lambda$CDM model defined by $(\xi_{\rm UV}, \, \xi_{\rm X}, \, T_{\rm min}, \, N_\alpha) = (50, \, 2 \times 10^{56} \, M_\odot, \, 5 \times 10^{4} \, K, \, 4000)$.

\begin{figure}
	\centering
	\includegraphics[width=0.49\textwidth]{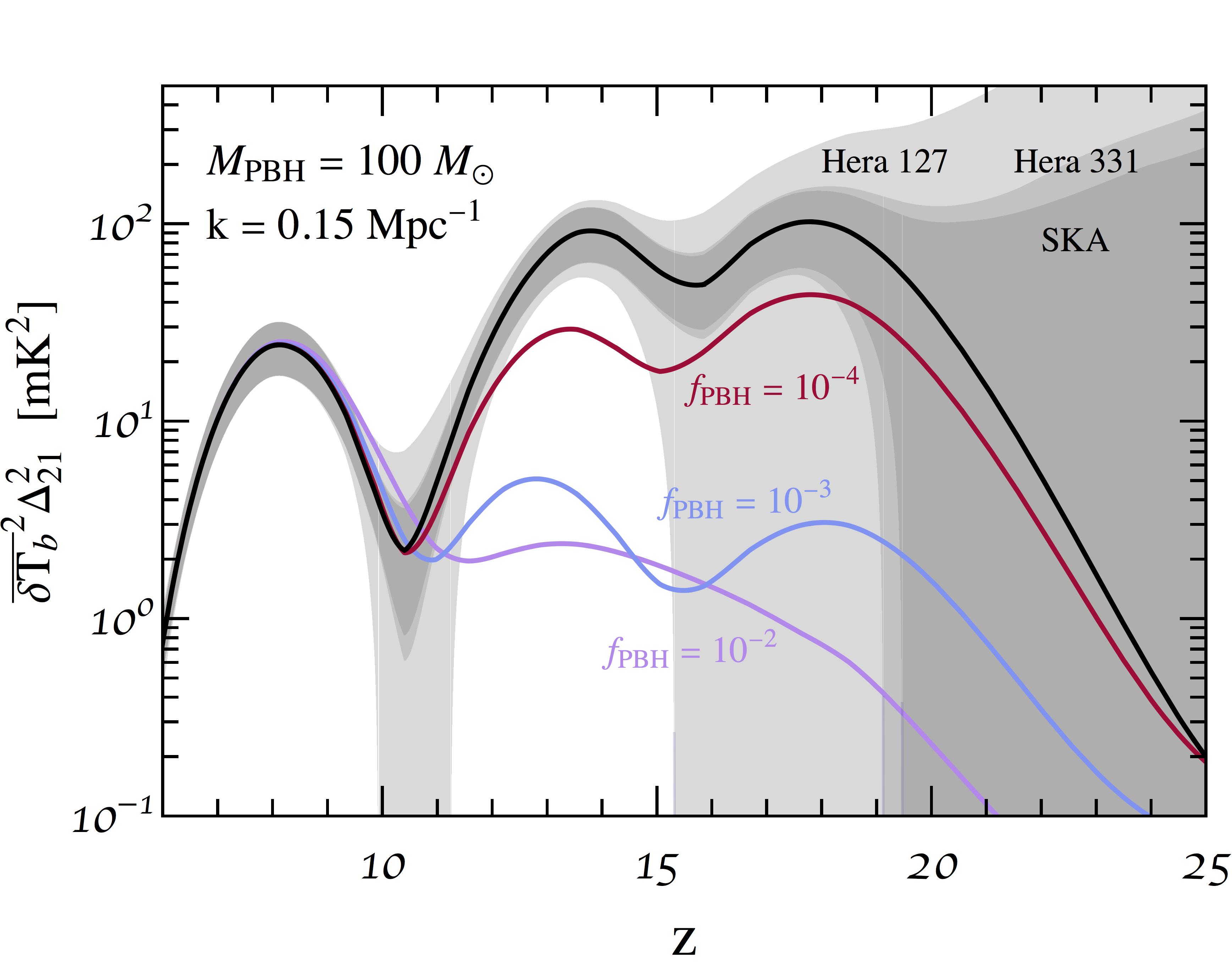}
	\includegraphics[width=0.49\textwidth]{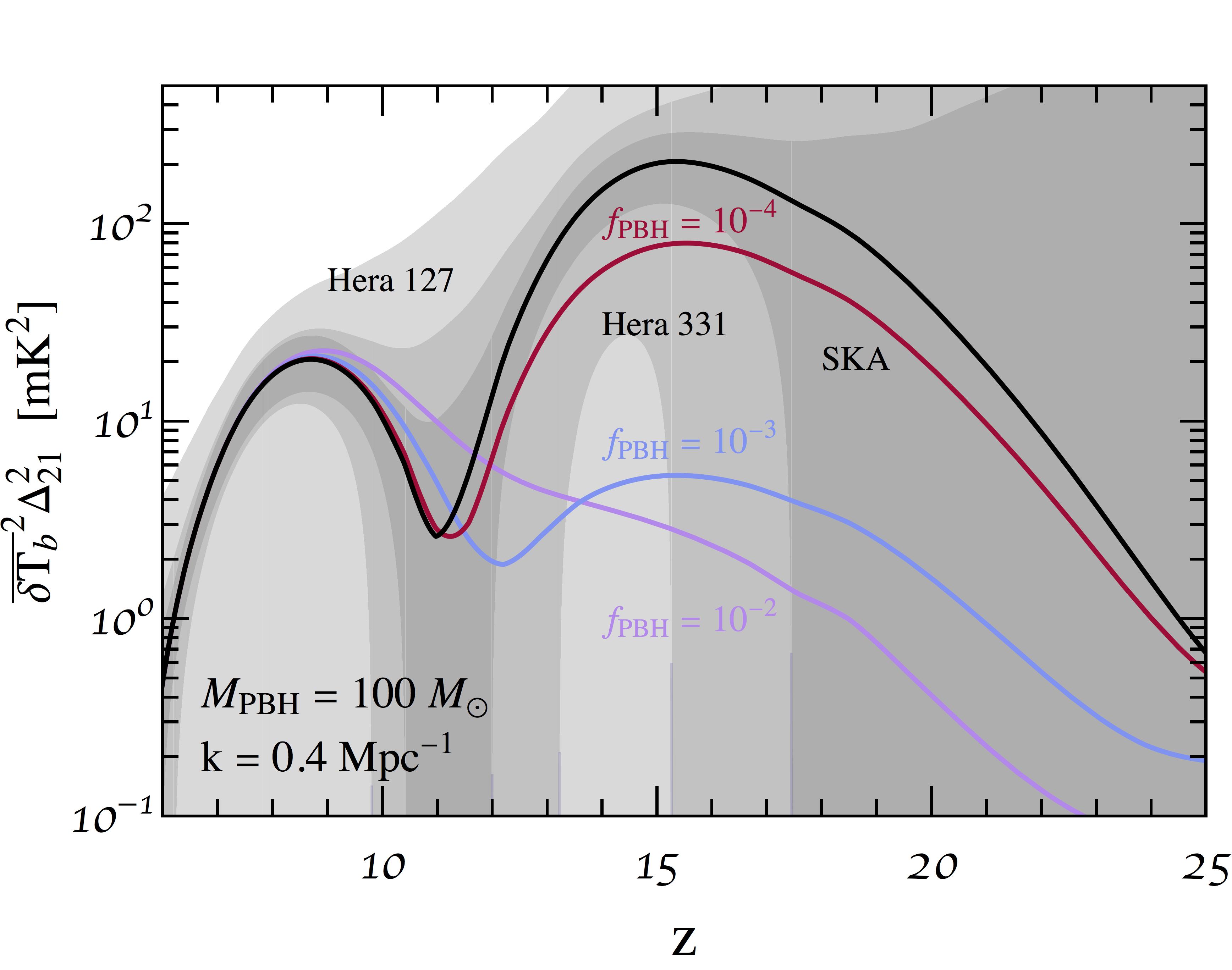}
	\caption{21cm power spectrum as a function of redshift $z$, for two values of the scale: $k = 0.15$~Mpc$^{-1}$ (left panel) and $k = 0.4$~Mpc$^{-1}$ (right panel). We consider a monochromatic PBH mass distribution with $M_{\rm PBH} = 100 \, M_\odot$ and the fiducial astrophysical parameters. We illustrate primordial black hole fractions of $f_{\rm PBH} = 10^{-2},  10^{-3}$ and $10^{-4}$. Errors for the future HERA and SKA radio interferometers for the standard scenario (with $f_{\rm PBH}=0$, denoted by the solid black lines) are also depicted. See text for details.}
	\label{fig:powerz}
\end{figure}

\begin{figure}
	\centering
	\includegraphics[width=0.49\textwidth]{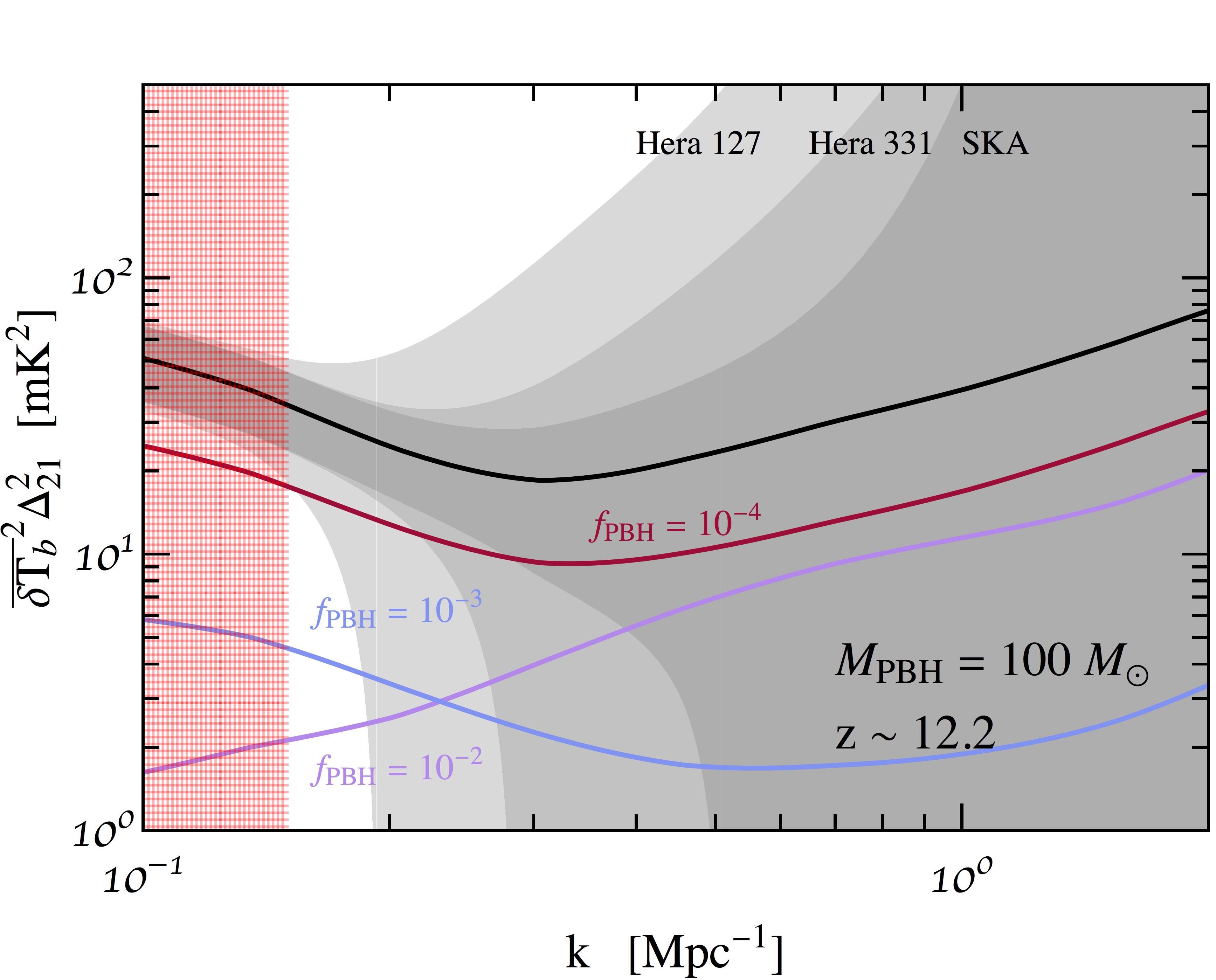}
	\includegraphics[width=0.49\textwidth]{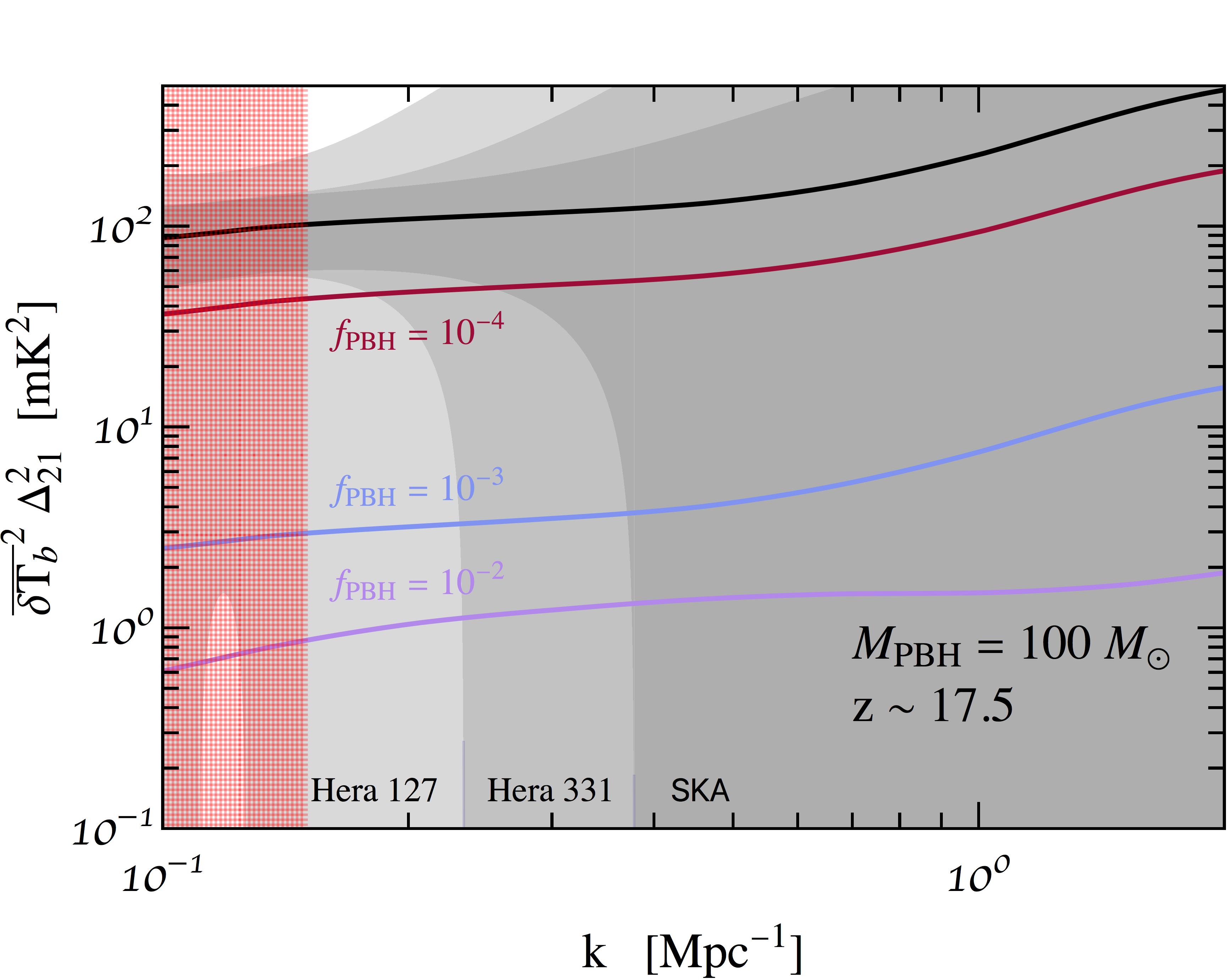}
	\caption{21cm power spectrum as a function of the scale, for two fixed redshifts: $z = 12.2$ (left panel) and $z = 17.5$ (right panel). The parameters are the same as in \Fig{fig:powerz}. The red areas denote the scales where foregrounds dominate over the signal.}
	\label{fig:powerk}
\end{figure}

In \Fig{fig:fig4} we showed the global differential 21cm brightness temperature for a population of $100 \, M_\odot$ PBHs, for a range of abundances, as a function of redshift. In \Fig{fig:powerz} we depict the 21cm power spectrum as a function of redshift at scales $k = 0.15$~Mpc$^{-1}$ (left panel) and $k = 0.4$~Mpc$^{-1}$ (right panel), which are expected to be reasonably free from foregrounds~\cite{Pober:2013ig}. Notice that the 21cm power spectrum in the standard scenario (solid black curves) exhibits three characteristic peaks, associated to the the epochs of reionization, heating from X-ray sources and Lyman-$\alpha$ pumping~\cite{Pritchard:2006sq, Mesinger:2010ne, Baek:2010cm}, from lower to higher redshifts. The presence of PBHs is translated into a suppression of the X-ray heating peak in the  power spectrum, which is obviously more pronounced as one increases $f_{\rm PBH}$. Additionally, the Lyman-$\alpha$ pumping peak could even disappear for $f_{\rm PBH} \gtrsim 10^{-2}$. In \Fig{fig:powerk} we show the 21cm power spectrum as a function of the scale for two fixed redshifts: $z = 12.2$ (left panel) and $z = 17.5$ (right panel). The red areas indicate the scales where the signal is expected to be contaminated by foregrounds. The effect of PBHs arises primarily from an increase in the flux of X-rays, which can induce both scale-dependent and redshift-dependent features in the power spectrum.

In Figs.~\ref{fig:powerz} and \ref{fig:powerk} we also include the forecasted errors associated to future HERA and SKA measurements of the 21cm power spectrum. These errors have been estimated by means of the publicly available code {\tt 21cmSense}\footnote{\url{https://github.com/jpober/21cmSense}}~\cite{Pober:2013jna, Pober:2012zz} (see also Ref.~\cite{Parsons:2011ew}). The total noise is given by
\begin{equation}
\overline{\delta \Delta^2_{T+S}}(k,z) = \left(\sum_i \frac{1}{\left(\Delta^2_{N,i}+\overline{\Delta}^2_{21}\right)^{2}}\right)^{-1/2} ~,
\end{equation}
with two contributions, one from thermal noise ($N$) plus a second one, a sample variance error ($S$), $\overline{\Delta}^2_{21} \equiv \overline{\delta T_b}^2 \Delta^2_{21}$. The thermal noise depends on the solid angle, the integrated observation time and the temperature of the system.  We focus here on the future HERA~\cite{Beardsley:2014bea} and SKA-low frequency~\cite{Mellema:2012ht} experiments. We  consider both the intermediate and final HERA configurations, with 127 and 331 antennas, which we refer to henceforth as Hera~127 and Hera~331. For SKA-low frequency, we follow the design presented in the SKA System Baseline Design Document~\cite{Mellema:2012ht}. Finally, we assume an exposure of 1080~hours and a bandwidth of 8~MHz, as these are the default values for these parameters in {\tt 21cmSense}.  

One of the fundamental difficulties associated with full statistical 21cm analyses arises from the fact that consistent and accurate theoretical predictions for the 21cm signal require time consuming semi-analytical (or even fully hydrodynamical) simulations. Restricting our attention momentarily to a single PBH mass, obtaining a coherent interpolation over a five dimensional parameter space (defined by $f_{\rm PBH}, \, \zeta_{\rm UV}, \, \zeta_{\rm X}, \, T_{\rm min}$, and $N_\alpha$) requires enormous amounts of computing time. Recently, various groups have developed techniques which rely on simplified calculations (at the expense of numerical accuracy), principle component decompositions, and machine learning algorithms, to reduce the computational expense involved in obtaining a comprehensive 21cm parameter scan (see, e.g., Refs.~\cite{Kern:2017ccn, Shimabukuro:2017jdh, Gillet:2018fgb, LaPlante:2018pst, Jennings:2018eko, Doussot:2019rdm}).

\begin{table}
	\setlength\extrarowheight{5pt}
	\begin{center}
		\begin{tabular*}{0.6\textwidth}{c @{\extracolsep{\fill}} c c c}
			\hline
			& Minimum value & Maximum value & Prior type \\ 
			\hline\hline
			$f_{\rm PBH}$ & $10^{-8}$ & 1 & Log \\ 
			\hline
			$\zeta_{\rm UV}$ & 10 & 100 & Log \\ 
			\hline
			$\zeta_{\rm X}$ & $2\times 10^{55} \, M_{\odot}^{-1}$ & $2\times 10^{57} \, M_{\odot}^{-1}$ & Log \\ 
			\hline
			$T_{\rm min}$ & $10^4 $ K & $10^5$ K & Log \\ 
			\hline
			$N_{\alpha}$ & $4\times 10^2$ & $4\times 10^4$ & Log \\ 
			\hline
		\end{tabular*}
	\end{center}
	\caption{Model parameters and priors adopted in the sensitivity analysis.}
	\label{table:priors} 
\end{table}

In this work, we circumvent this numerical issue by exploiting a class of feed-forward neural networks known as multilayer perceptrons (MLPs). Specifically, for a fixed PBH mass, we compute the full 21cm power spectrum for $\sim 600$ choices of astrophysical parameters (constrained to the range defined in \Tab{table:priors}), and construct MLPs with two hidden layers, each containing $\sim 50$ hidden nodes, to emulate the output of {\tt 21cmFAST} for arbitrary choices of parameters. The MLP is trained on $\sim70\%$ of the computed power spectra while the remaining $30\%$ are simultaneously used to ensure the neural network is not over-learning. We find that this procedure reproduces the computed datasets (both the trained dataset and test dataset), as well as various power spectra computed with randomly generated points in parameter space after the training of the neural network. We choose to assess the relative accuracy of the neural net by defining an accuracy statistic, which later serves as the quantity directly entering our likelihood, as
\begin{equation}
\label{eq:alphaErr}
\alpha_{\rm err} \equiv \left|\frac{\bar{\Delta}^2_{21}(k,z) - {\Delta}^2_{21, NN}(k,z)}{\sqrt{ \overline{\delta\Delta^2_{T+S}}^{\, 2} (k,z)+ (0.4 \times \bar{\Delta}^2_{21}(k,z))^2}}\right| ~,
\end{equation}
where $\bar{\Delta}^2_{21}(k,z)$ is the power spectrum computed by {\tt 21cmFAST}, and ${\Delta}^2_{21, NN}(k,z)$ is the power spectrum as computed by the neural net. Following Ref.~\cite{Greig:2015qca}, we have chosen to add in quadrature an additional modeling error (here we conservatively adopt a $40\%$ error, although we note that this factor is somewhat {\emph{adhoc}}) that is intended to capture the approximate level of disagreement between {\tt 21cmFAST} calculations and full hydrodynamic simulations. Using \Eq{eq:alphaErr}, one finds that the mean error of the neural net is $\lesssim 1\%$.

\pagebreak
\subsection{Statistics}
\label{sec:stats}

We adopt a multivariate Gaussian likelihood with a diagonal covariance matrix, the elements of which are set using \Eq{eq:alphaErr}, and further assume that measurements are obtained in six log-spaced bins in $k-$space from $k = 0.15~{\rm Mpc}^{-1}$ to $k =1~{\rm Mpc}^{-1}$ and 9 log-spaced measurements in redshift from $z \sim 8.3$ to $z \sim 19.5$. While it is likely that experiments such as SKA will be capable of achieving far better resolution in both $k-$ and $z-$space, adopting a finer grid without properly accounting for correlated errors risks the possibility of significantly overestimating the sensitivity of these experiments.

We adopt the four-parameter astrophysical model discussed in \Sec{subsec:numer} (implying a total of 5 model parameters for each fixed value of $M_{\rm PBH}$), where the priors on each parameter are as given in \Tab{table:priors}. For each fixed value of $M_{\rm PBH}$, we perform a Markov chain Monte Carlo (MCMC) assuming the true measurement is given by a fiducial astrophysical model with $f_{\rm PBH} = 0$.

\begin{figure}
	\centering
	\includegraphics[width=\textwidth]{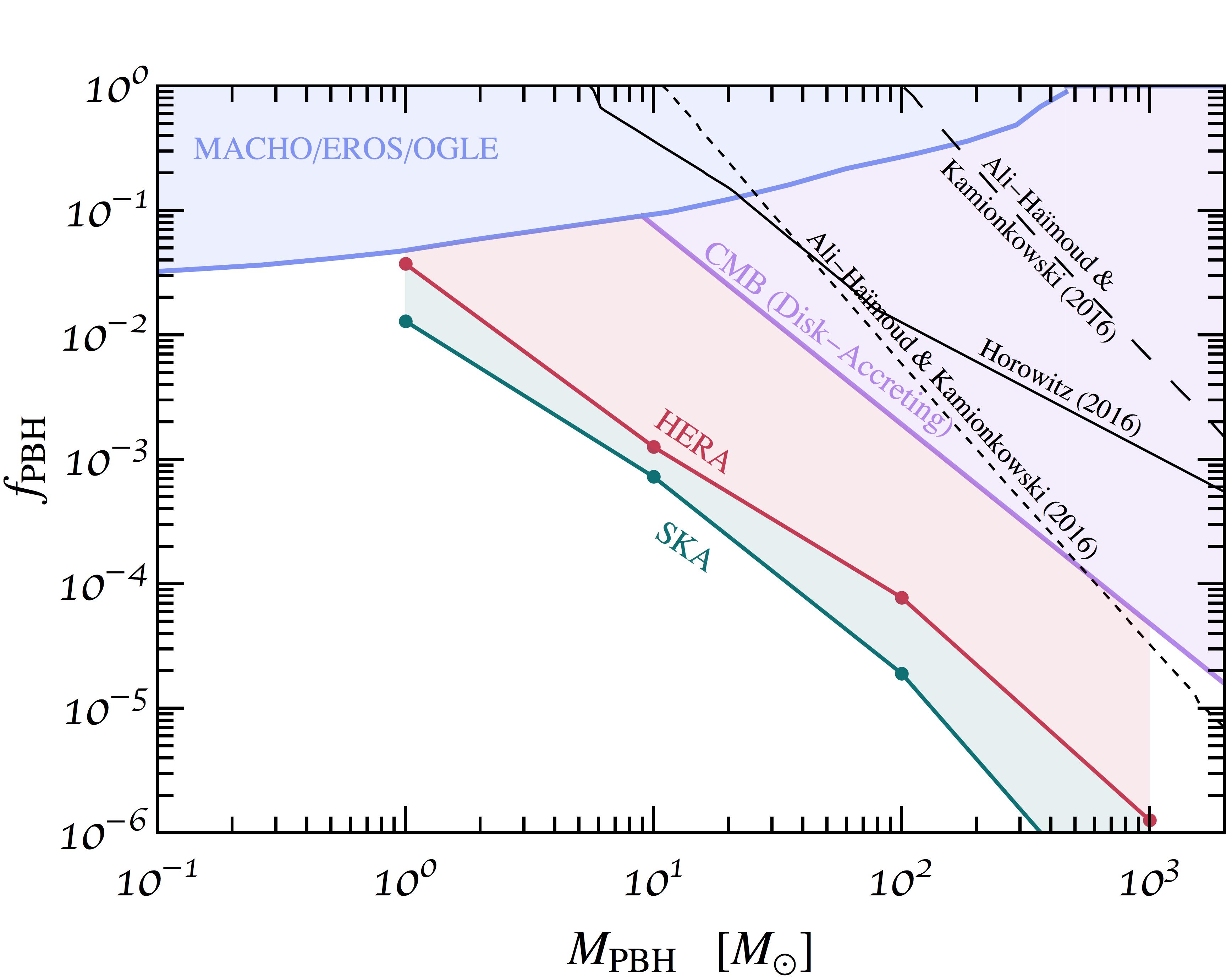}
	\caption{Estimated $2\sigma$ sensitivity of SKA (green line) and HERA with 331 antennas (red line) for a monochromatic distribution of PBHs in the mass range $M_\odot \leq M_{\rm PBH} \leq 10^3 M_\odot$. Results are compared to various limits derived from microlensing surveys~\cite{Allsman:2000kg, Tisserand:2006zx, Wyrzykowski:2011tr} (blue line) and the CMB~\cite{Ali-Haimoud:2016mbv, Horowitz:2016lib, Poulin:2017bwe} (purple and black lines).}
	\label{fig:sense}
\end{figure}

\begin{figure}
	\centering
	\includegraphics[width=\textwidth]{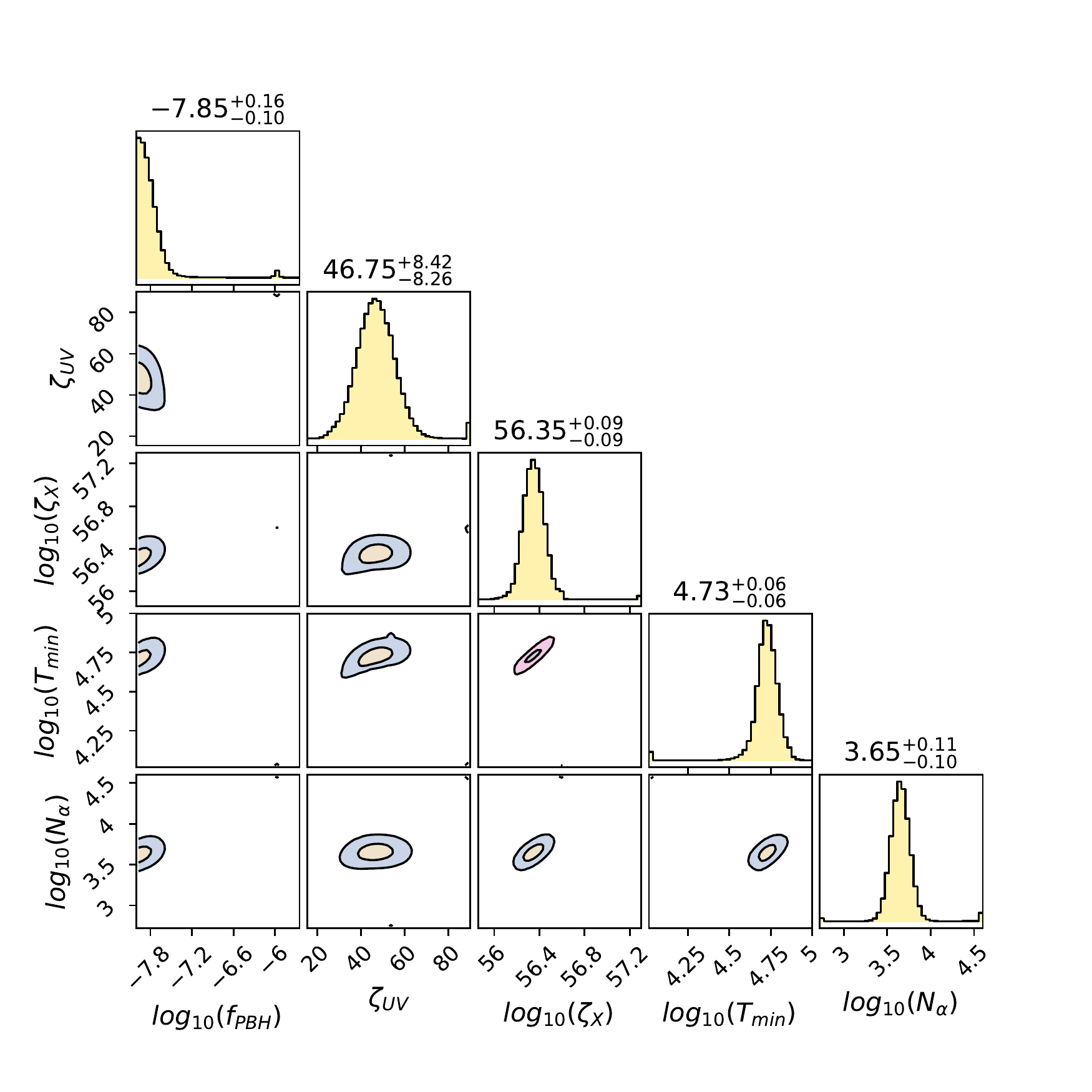} \vspace{-6mm} 
	\caption{Parameter sensitivity for SKA MCMC, assuming a monochromatic PBH mass distribution with $M_{\rm PBH} = 10^3 \, M_\odot$, depicted with $1\sigma$ and $2\sigma$ contours. The numbers above each panel denote the maximum of the posterior and the upper/lower $1\sigma$ confidence interval on these values.}
	\label{fig:triangle}
\end{figure}

\section{Results and conclusions}
\label{sec:conclusions}

The $2\sigma$ upper limits as derived from the MCMC analysis described in \Sec{sec:stats} are shown in \Fig{fig:sense} for HERA (red line), assuming 331 antennas, and SKA (green line). Also shown in \Fig{fig:sense} are current constraints from microlensing surveys~\cite{Allsman:2000kg, Tisserand:2006zx, Wyrzykowski:2011tr} (blue line), and the CMB~\cite{Ali-Haimoud:2016mbv, Horowitz:2016lib, Poulin:2017bwe} (purple and black lines), where the purple contour denotes the bound derived assuming an identical accretion model to the one adopted here. As discussed in the introduction, there exist additional constraints in this region, however these are not shown for clarity. The MCMC analysis was only performed for PBH masses in the range $(1 - 10^3)~M_\odot$, however this is a consequence of the computational difficulty associated with 21cm analyses. In principle, these experiments will be sensitive to both smaller and larger masses, and the relative importance of the shot noise, minihalos, and the modification to the IGM, in general, varies with mass. 

The results shown in \Fig{fig:sense} suggest that near-future 21cm experiments will be able to increase the sensitivity to $\mathcal{O}(M_\odot)$ PBHs relative to that of the CMB by more than one order of magnitude. Moreover, 21cm surveys provide a highly complementary probe that is sensitive to the dynamics of accretion at much lower redshifts. Thus, should a positive detection be made with both probes, one may be able to shed light on the dynamics and evolution of accretion in these systems. For completeness, we also show in \Fig{fig:triangle} the result of the MCMC performed for $M_{\rm PBH} = 10^3 \, M_\odot$ using the SKA telescope array. \Fig{fig:triangle} illustrates that many of the astrophysical parameters adopted in this analysis could be well constrained by these experiments.

In this work we have focused on understanding the sensitivity that near-future radio interferometers may have to a population of $\gtrsim \mathcal{O}(M_\odot)$ PBHs using the 21cm power spectrum. In particular, we have jointly analyzed three effects: namely, heating and ionization arising from accretion on both local and global scales, and the signal arising from minihalos (whose number density is enhanced with respect to that of $\Lambda$CDM, a consequence of having non-negligible Poissonian noise in the distribution of PBHs). Contrary to previous claims in the literature~\cite{Tashiro:2012qe, Bernal:2017nec}, we find that the local effect of accretion is negligible for all redshifts, masses, and PBH fractions discussed in this work. Moreover, for the PBH masses studied here, the addition of shot noise to the power spectrum modifies the halo mass function below the mass threshold for efficient star formation. Consequently, this term would only be relevant if minihalos gave sizable contributions, or if one considers extremely heavy PBHs (however it is worth noting that there unavoidably exists uncertainty in the high redshift halo mass function within $\Lambda$CDM which could produce degenerate effects to minihalos). While previous work has claimed that the enhanced number density of minihalos, produced by viable populations of PBHs, could give rise to an observable signal~\cite{Gong:2017sie, Gong:2018sos}, we show here that these calculations had not self-consistently accounted for the global heating of the IGM, which significantly raises the Jeans mass and tends to suppress the minihalo contribution. Thus, the dominant effect on the 21cm power spectrum arises exclusively from the modifications to the heating and ionization of the IGM. We find experiments like HERA and SKA could significantly improve upon existing limits from the CMB, should observations be inconsistent with a $f_{\rm PBH} = 0$ cosmology (i.e., no PBHs).

Finally, we emphasize that 21cm observations will have access to much more information than just the two-point correlation function. Should foregrounds be removed to a sufficiently high degree, these experiments may be able to exploit higher dimensional n-point correlation functions, or perhaps one could exploit the power of convolutional neural nets to shed light of the highly non-Gaussian behavior of neutral hydrogen during these epochs.

\section*{Acknowledgments}
OM, PVD and SW are supported by the Spanish grant FPA2017-85985-P of the MINECO.
SPR is supported by a Ram\'on y Cajal contract, by the Spanish MINECO under grant FPA2017-84543-P, and partially, by the Portuguese FCT through the CFTP-FCT Unit 777 (UID/FIS/00777/2019).
All the authors also acknowledge support from the European Union's Horizon 2020 research and innovation program under the Marie Sk\l odowska-Curie grant agreements No.\ 690575 and 674896. 
This work was also supported by the Spanish MINECO grant SEV-2014-0398.

\bibliography{biblio}

\end{document}